\documentclass[useAMS,usenatbib]{mn2e}
\usepackage[utf8]{inputenc}
\usepackage{rotating}
\usepackage{booktabs}
\usepackage{setspace}
\usepackage{txfonts}
\usepackage{float}
\usepackage{amssymb}
\usepackage{epsfig}
\usepackage{color}
\usepackage{graphicx}
\usepackage{multirow}
\usepackage{longtable}
\usepackage{verbatim}
\usepackage[flushleft]{threeparttable}
\usepackage{gensymb}
\usepackage[export]{adjustbox}
\usepackage{subfig}

\usepackage{array}
\usepackage{natbib}

\newcommand{\rx}{\textsc{RXTE}}
\newcommand{\sub}{4U~1728--34}
\newcommand{\xmm}{\textsc{XMM-Newton}}
\newcommand{\Nu}{\textsc{NuSTAR}}
\newcommand{\sw}{\textsc{Swift}}

\title[The X-ray properties of 4U~1728--34]
{
Study of the X-ray properties of the neutron-star binary 4U~1728--34 from the soft to hard state
}

\author[Yanan Wang et al.]{Yanan Wang$^{1}$\thanks{E-mail: yanan@astro.rug.nl}, Mariano M\'endez$^{1}$, Diego Altamirano$^{2}$, Guobao Zhang$^{3,4}$, T. M. Belloni$^{5}$,  \newauthor
Evandro M. Ribeiro$^{1}$, M.~Linares$^{6,7}$, Andrea Sanna$^{8}$, S.E. Motta$^9$, John A. Tomsick$^{10}$\\
$^{1}$Kapteyn Astronomical Institute, University of Groningen, PO BOX 800, NL-9700 AV Groningen, the Netherlands\\
$^{2}$Physics \& Astronomy, University of Southampton, Southampton, Hampshire SO17 1BJ, UK\\
$^3$ Key Laboratory for the Structure and Evolution of Celestial Objects, Chinese Academy of Sciences, 650216 Kunming, P. R. China\\
$^4$ Center for Astronomical Mega-Science, Chinese Academy of Sciences, 20A Datun Road, Chaoyang District, Beijing, 100012, P. R. China\\
$^{5}$Istituto Nazionale di Astrofisica, Osservatorio Astronomico di Brera, Via E. Bianchi 46, I-23807 Merate, Italy\\
$^{6}$Departament de F{\'i}sica, EEBE, Universitat Polit{\`e}cnica de Catalunya, Av. Eduard Maristany 16, E-08019 Barcelona, Spain\\
$^{7}$Institute of Space Studies of Catalonia (IEEC), E-08034 Barcelona, Spain\\
$^{8}$Dipartimento di Fisica, Universit\'{a} degli Studi di Cagliari, SP Monserrato-Sestu km 0.7, I-09042, Monserrato, Italy\\
$^{9}$University of Oxford, Department of Physics, Astrophysics Denys Wilkinson Building Keble Road, Oxford OX1 3RH, UK\\
$^{10}$Space Sciences Laboratory, 7 Gauss Way, University of California, Berkeley, CA 94720-7450, USA \\}

\begin{document}

\date{Accepted ? December ?. Received ? December ?; in original form ? December ?}
\maketitle

\begin{abstract}
We studied five {\xmm} observations of the neutron-star binary {\sub} covering the hard, intermediate and soft spectral states. 
By jointly fitting the spectra with several reflection models, we obtained an inclination angle of $25\degree-53\degree$ and an iron abundance up to 10 times the solar. 
From the fits with reflection models, we found that the fluxes of the reflection and the Comptonised components vary inconsistently; since the latter is assumed to be the illuminating source, this result possibly indicates the contribution of the neutron star surface/boundary layer to the disc reflection. 
As the source evolved from the relatively soft to the intermediate state, the disc inner radius decreased, opposite to the prediction of the standard accretion disc model.
We also explore the possible reasons why the supersolar iron abundance is required by the data and found that this high value is probably caused by the absence of the hard photons in the {\xmm} data.
\end{abstract}


\section{Introduction} \label{intro}
A reflection spectrum, as the result of the hard coronal radiation illuminating an accretion disc, has been observed in several accreting black hole (BH, e.g., \citealt{George1991,Magdziarz1995,Nowak2002,Miller2013}) and neutron star (NS, e.g., \citealt{Bhattacharyya2007,Cackett2008,Cackett2010,Wang2017}) systems.
The combination of the high fluorescent yield and large cosmic abundance makes the iron emission line at 6.4--7~keV the most prominent feature in the reflection spectrum of these systems (see the Monte-Carlo simulation results in \citealt{Reynolds1996}). As the energy of some incident X-ray photons is much larger than the binding energy of the atomic electron in the disc, where those photons are scattered, Compton recoil occurs.
This leads to a hump at high energies (e.g., \citealt{Matt1991,George1991}), known as the Compton hump, peaking at $30$~keV in the reflection spectrum. 

Unlike in BH and faint NS low-mass X-ray binaries (LMXBs), where the illuminating source of the disc is assumed to be a hot corona of highly energetic elements, in bright NS-LMXBs the NS boundary layer could contribute significantly to the reflection spectrum as well \citep{Cackett2010,Miller2013,Ludlam2017}.
Regardless of the nature of the illuminating source, when reflection occurs in the vicinity of the compact object, the reflection spectrum can be modified by Doppler effects, light bending, and gravitational redshift; the combination of all these effects produce a broadened and skewed line profile with a red wing extending to low energies (e.g., \citealt{Fabian2000,Reynolds2003,Miller2008}).
Therefore, by studying the asymmetrically broadened profile of such lines, 
we can investigate the geometry and the extension of the accretion disc.

Even though modelling the reflection spectrum has so far provided one of the best methods to estimate the spin parameter in BH systems, the derived high iron abundance (several times the solar value, e.g., Cyg~X--1, \citealt{Parker2015}; GX~339--4, \citealt{Furst2015,Garcia2015}) of the disc rises concerns about the accuracy of the spin estimates. Currently there is no plausible physical explanation for these systems to be so iron rich. \cite{Furst2015} found that the high iron abundance in GX~339--4 is model-dependent. Once they allowed the photon indices of the direct power-law component and the power-law component that illuminates the disc to be different, the best-fitting iron abundance decreased and the fit statistically improved. Alternatively, \cite{Tomsick2018} reported that by applying high density (up to $10^{22}~\rm cm^{-3}$) reflection models, the fit no longer required a supersolar iron abundance in Cyg~X--1. 

{\sub} is a weakly magnetized neutron star accreting from a hydrogen-poor donor star \citep{Shaposhnikov2003,Galloway2010}. It has been classified as an ultra-compact, atoll-type, LMXB with high Galactic hydrogen column density, $N\rm_{H} = 2.4-4.5 \times 10^{22}~cm^{-2}$ \citep{Dai2006,Egron2011,Sleator2016,Mondal2017}.
Type I bursts and burst oscillations at $\sim363$~Hz have been reported in several works for this source (e.g., \citealt{Strohmayer1996,Zhang2016,Verdhan2017}). 
The distance to {\sub} has been estimated to be in the range 4.4--5.1~kpc using the Eddington limit luminosity of the photospheric radius expansion bursts \citep{Di2000,Galloway2003}.
Kilohertz quasi-periodic oscillations (kHz~QPOs) have been detected in the persistent emission (e.g., \citealt{Strohmayer1996,Migliari2003,Mukherjee2012,Verdhan2017}).

The source states in atoll-type NSs are called the `island' and `banana' states, based on the shape of the colour-colour diagram (CD) and the timing properties of these sources, which correspond to the `hard' and `soft' states in other X-ray binaries, respectively. 
We used the latter nomenclature hereafter in this paper.
The source states in these systems, which is likely related to changes in the mass accretion rate \citep{Hasinger1989}, are usually associated with the evolution of the accretion flow. 
For instance, as a source evolves from the soft to the hard state, the edge of the disc moves outwards, from the innermost stable circular orbit (ISCO) to a larger radius (e.g., \citealt{Esin1997,Done2007}). 
However, Sanna et al. (2014) found that the inner radius of the accretion disc was uncorrelated with the spectral state for the neutron star 4U~1636--53.


A broad iron emission line has been detected in the X-ray spectra of {\sub} with several instruments, e.g., \textsc{BeppoSAX} \citep{Di2000,Piraino2000}, {\xmm} \citep{Ng2010,Egron2011}, \textsc{AstroSat}/\textsc{LAXPC} \citep{Verdhan2017}, \textsc{NuSTAR} and \textsc{Swfit} \citep{Sleator2016,Mondal2017}.
Both \cite{Sleator2016} and \cite{Mondal2017} fitted the spectra of the {\sw} and {\Nu} data of this source using a reflection model.
\cite{Sleator2016} found a disc inclination angle of $\sim 37\degree$, an iron abundance of the accretion disc of $< 1$ times solar and an upper limit for the inner disc radius of $\leq2~R_{g}$, where $R_{g}=GM/c^{2}$.
\cite{Mondal2017} reported that the inclination angle in this system is $22\degree-40\degree$, the disc iron abundance is $2-5$ times solar and, as the source evolved from the soft to the hard state, the inner radius changed from $2.3_{-1.0}^{+2.1}$ to $3.7_{-0.7}^{+2.2}$~$R\rm_{ISCO}$, consistent with being constant.

In this paper, we conduct timing and spectral analysis of the neutron-star LMXB {\sub} with {\xmm} data and the (quasi) simultaneous {\rx} data to study how the accretion flow changed while the source evolved from the soft to hard state. The paper is organized as follows: in Section~2 we describe the observations and the data reduction; our results of the spectral analysis are presented in Section~3; we discuss our results in Section~4, and we summarize our conclusions in Section~5.

\section{Observations and data reduction}
The \textsc{XMM-Newton} observatory \citep{Jansen2001} carries 3 high throughput X-ray telescopes, each of them containing an European Photon Imaging Camera (EPIC, 0.1$-$12~keV).
Two of these cameras are equipped with Metal Oxide Semi-conductor (MOS) CCDs \citep{Turner2001} and one carries \textsc{PN} CCDs \citep{Strder2001}.
Reflection grating spectrometers (RGS, 0.35--2.5~keV, \citealt{Herder2001}) are installed behind two of these telescopes.

The five {\xmm} observations of {\sub} used here were taken between August 28 and October 7, 2011. We show the details of the observations in Table~\ref{tab:1} and refer to them as Obs.~1 to 5 according to the observing time.
We used data obtained with the EPIC-\textsc{PN} in Timing mode and with the RGS in Standard spectroscopy.
To reduce and analyse the raw data we used version 16.1.0 of the {\xmm} Scientific Analysis Software (SAS) package. 
Using the command {\it epatplot}, we found that the \textsc{PN} data were affected by pile up and we hence excluded the central region of the point-spread function source to mitigate this effect.

There were 14 type-I X-ray bursts in the \textsc{PN} light-curves; we excluded these periods when we produced the \textsc{PN} spectra.
We extracted all the \textsc{PN} background spectra from the outer columns of
the central CCD (RAWX in 4--10) and found that the extracted background spectra are contaminated with the source (see also \citealt{Ng2010,Hiemstra2011}). We hence used the \textsc{PN} observation (ObsID~0085680601) of GX~339--4, which is on similar sky coordinates and column density along the line of sight, when this source was in the quiescent state, as a blank field to extract background spectra for all the five \textsc{PN} observations. 
We re-binned the \textsc{PN} spectra to have a minimum of 25 counts or to oversample the instrumental energetic resolution by a maximum factor of 3 in each bin.
We fitted the \textsc{PN} spectra between 2.5 and 11~keV, avoiding the detector Si K-edge at 1.8~keV and the mirror Au M-edge at 2.3~keV \citep{Egron2011}.


We extracted the RGS data using the SAS tool {\it rgsproc} 
to produce calibrated event files, spectra and response matrices.
The RGS data were grouped to provide a minimum of 25 counts per bin.
We fitted the RGS spectra between 1 and 2~keV to constrain models in the soft band.
We fitted the X-ray spectra using XSPEC (12.9.1a).
To account for the interstellar absorption, in all fits we used the component {\sc tbabs} with solar abundances from \cite{Wilms2000} and cross-sections from \cite{Verner1996}. Unless explicitly mentioned, we quote all errors at $1\sigma$ confidence level and at 95\% confidence for upper limits.

There were also 22 Rossi X-ray Timing Explorer ({\rx}) observations of 4U~1728--34 (quasi) simultaneous with our {\xmm} data. 
To search for the presence of QPOs, we first generated standard \textit{good-time interval} files (GTIs) to remove instrumental drop-outs and other technical anomalies from the Proportional Counting Array (PCA) observations as suggested by the {\rx} Documentation\footnote{https://heasarc.gsfc.nasa.gov/docs/xte/abc/screening.html}. Type I bursts have been detected and removed as well.
We then divided each observation into segments of 16 seconds and extracted power spectra using the full energy band with a Nyquist frequency of 2048 Hz and averaged all the segments to obtain a single power spectrum for each observation.

\begin{table}
\caption{\label{tab:1}{\xmm} observations of 4U 1728--34 used in this paper}
\renewcommand{\arraystretch}{1.2}
\setlength{\tabcolsep}{4pt}
\footnotesize
\centering
\begin{tabular}{cccccc}
\hline \hline
Obs.&ObsID&Instrument&Start Date& Exposure (ks)& $S_a$\\

\hline
1&0671180201& \textsc{PN} &2011-08-28&52.1~(52.0$^*$)&2.3\\
 &         & RGS 1/2&                           &53.4\\
2&0671180301& PN &2011-09-05&46.7~(46.7$^*$)&1.8\\
  &        & RGS 1/2&                           &51.9  \\       
3&0671180401& PN &2011-09-17&52.4~(52.2$^*$)&1.6\\
  &        & RGS 1/2&                           &54.0\\ 
4&0671180501& PN &2011-09-27&50.6~(50.5$^*$)&1.5\\
  &        & RGS 1/2&                           &51.9\\
5&0671180601& PN &2011-10-06&57.6~(57.4$^*$)&1.3\\
    &     & RGS 1/2&                           &58.9\\ 
\hline
\end{tabular}
\begin{flushleft}
$^*$Final exposure time excluding bursts.\\
\end{flushleft}
\end{table}

\section{Results}
\subsection{Timing analysis \label{sec:qpo}}
According to \cite{Zhang2016}, some of the {\rx} observations of {\sub} are contaminated by the nearby active transient 4U~1730--335 (the Rapid Burster). Both of the sources are in the PCA field of view and this transient was in outburst at the same time with the {\rx} observations. 
Because the Rapid Burster only displayed significant power in the low-frequency range \citep{Rutledge1995,Stella1988}, we ignored the frequency range, $<200$~Hz, of the power spectra of {\sub} to avoid the contamination from the Rapid Burster.
We linearly rebinned the power spectra by a factor of 200 to a frequency resolution of 12.5~Hz to improve the signal-to-noise ratio and fitted the power spectra with a constant to represent the Poisson noise and one or two Lorentzians to represent the kHz QPO(s).

We found significant kHz QPOs only in two observations: ObsIDs 96322-01-03-00 and 96322-01-03-01, both of them corresponding to Obs.~3 of the {\xmm} data. 
The kHz QPO in observation 96322-01-03-00 has a frequency of $604\pm17$~Hz and a fractional rms amplitude of $7.3\pm1.8$\%, at a level of significance of $2.8~\sigma$, calculated as the ratio of the integral power of the fitted Lorentzian with $1~\sigma$ negative error. Another kHz QPO in observation 96322-01-03-01 has a frequency of $583\pm19$~Hz and a fractional rms amplitude of $9.8\pm1.6$\%, at a level of significance of $4~\sigma$.


\subsubsection{Colour-colour diagram and long-term light curve}
\begin{figure} 
\centering  
\resizebox{1\columnwidth}{!}{\rotatebox{0}{\includegraphics[clip]{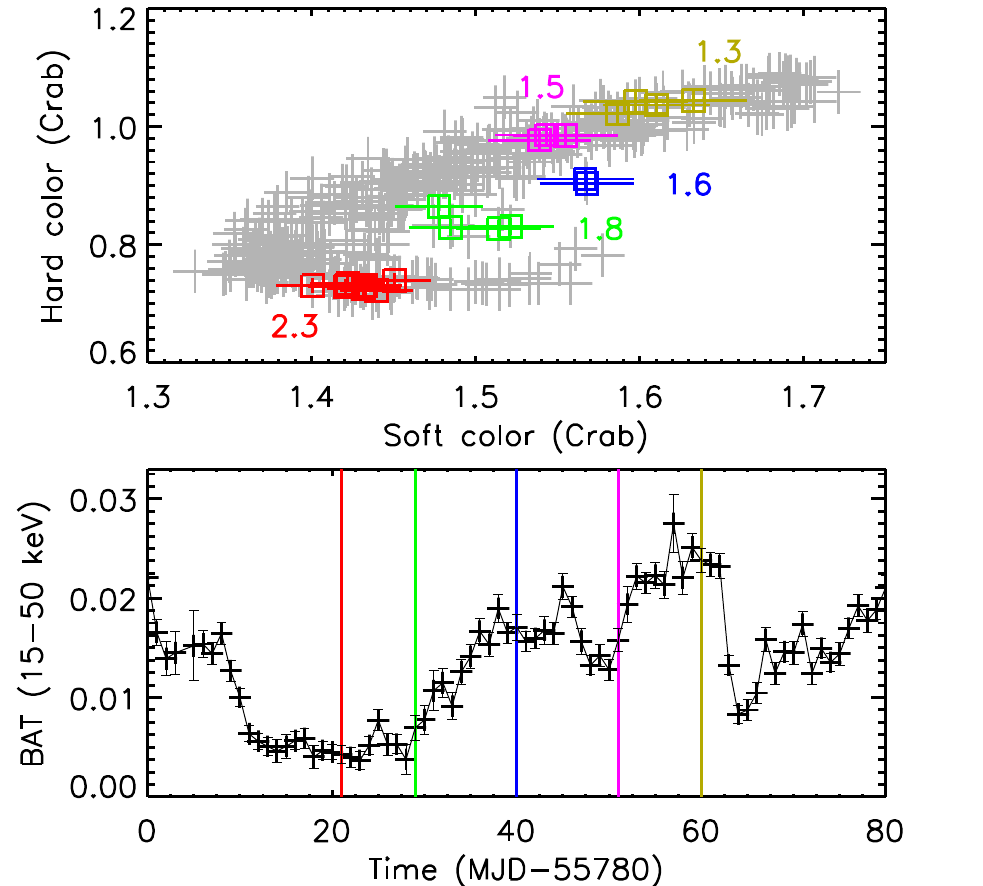}}}
\vspace{-5mm}
\caption{Upper panel: {\rx} colour-colour diagram of {\sub}. Each point corresponds to one {\rx} observation. The numbers represent the values of $S_a$ of each {\xmm} observation, respectively.
Lower panel: {\sw}/BAT ($\rm cts~s^{-1}~cm^{-2}$ in 15--50~keV) long-term light curve of {\sub}. Each point corresponds to one-day {\sw} observation.
The {\xmm} observations listed in Table~1 from top to bottom correspond to the simultaneous {\rx}/{\sw} data in red, green, blue, magenta and olive squares/lines.}
\label{fig:ccd}
\end{figure}

To explore the source state of {\sub}, we took the data from \cite{Zhang2016} and plotted the CD of the {\rx} data in the upper panel of Fig.~\ref{fig:ccd}.
As the definition of the colours in their work, the soft and hard colours are the 3.5$-$6.0/2.0$-$3.5~keV and 9.7$-$16.0/6.0$-$9.7~keV count rate ratios, respectively. 
Type I bursts have been removed from the {\rx} data and the colours of {\sub} are normalized to the colours of Crab. 
\cite{Zhang2016} parametrized the position of the source on the CD through the value of the parameter $S_{a}$, that gives quantitatively the position of the source along the path traced by the source in the CD \citep{Mendez1999a}. 
They fixed the values of $S_a=1$ and $S_a=2$ at the top-right and the bottom-left vertex of the CD, respectively.

We assigned an $S_{a}$ value to each {\xmm} observation as the average $S_{a}$ value of the simultaneous {\rx} data and indicated them with the red, green, blue, magenta and olive squares in the upper panel of Fig.~\ref{fig:ccd}. During our observations, the source evolved from the left bottom to the right top on the CD as $S_a$ decreased.
As some of the {\rx} observations are contaminated by the Rapid Burster, this prevented us from using the simultaneous {\rx} data to do spectral analysis and the emission from the Rapid Burster may have also affected the colours of these observations. For instance, Obs.~2 and 3 are off the main track in the CD and both of them are entirely contaminated; the other {\rx} observations are only partially contaminated.
To check whether the evolution of the source in the {\rx} CD is reliable, we created a {\sw}/BAT long-term light curve in the energy of 15--50~keV at around the time of the observations with {\xmm} of {\sub}; we show the {\sw}/BAT light curve in the lower panel of Fig.~\ref{fig:ccd}. Corresponding to the five {\xmm} observations, the count rate of the {\sw}/BAT light curve increased from Obs.~1 to 3, remained constant within errors during Obs.~3 and 4, and increased again from Obs.~4 to 5.

Both the source evolution on the CD in the upper panel of Fig.~\ref{fig:ccd} and of the light curve in the lower panel of Fig.~\ref{fig:ccd} indicate that the source indeed transited from the relatively soft to the hard state.

\subsection{Spectral analysis}
We initially fitted the \textsc{PN} spectra of the five {\xmm} observations in the energy range 2.5--11~keV with a thermally Comptonised component ({\sc nthcomp} in XSPEC, \citealt{Zdziarski1996,zycki1999}) plus a single temperature blackbody component ({\sc bbodyrad} in XSPEC). The fit was bad, $\chi^2=2199.9$ for $\nu=652$, where $\nu$ is the number of degrees of freedom (d.o.f.), and the fit showed prominent residuals at 5--8.5~keV.

\begin{figure}
\includegraphics[width=6.3cm,angle=270]{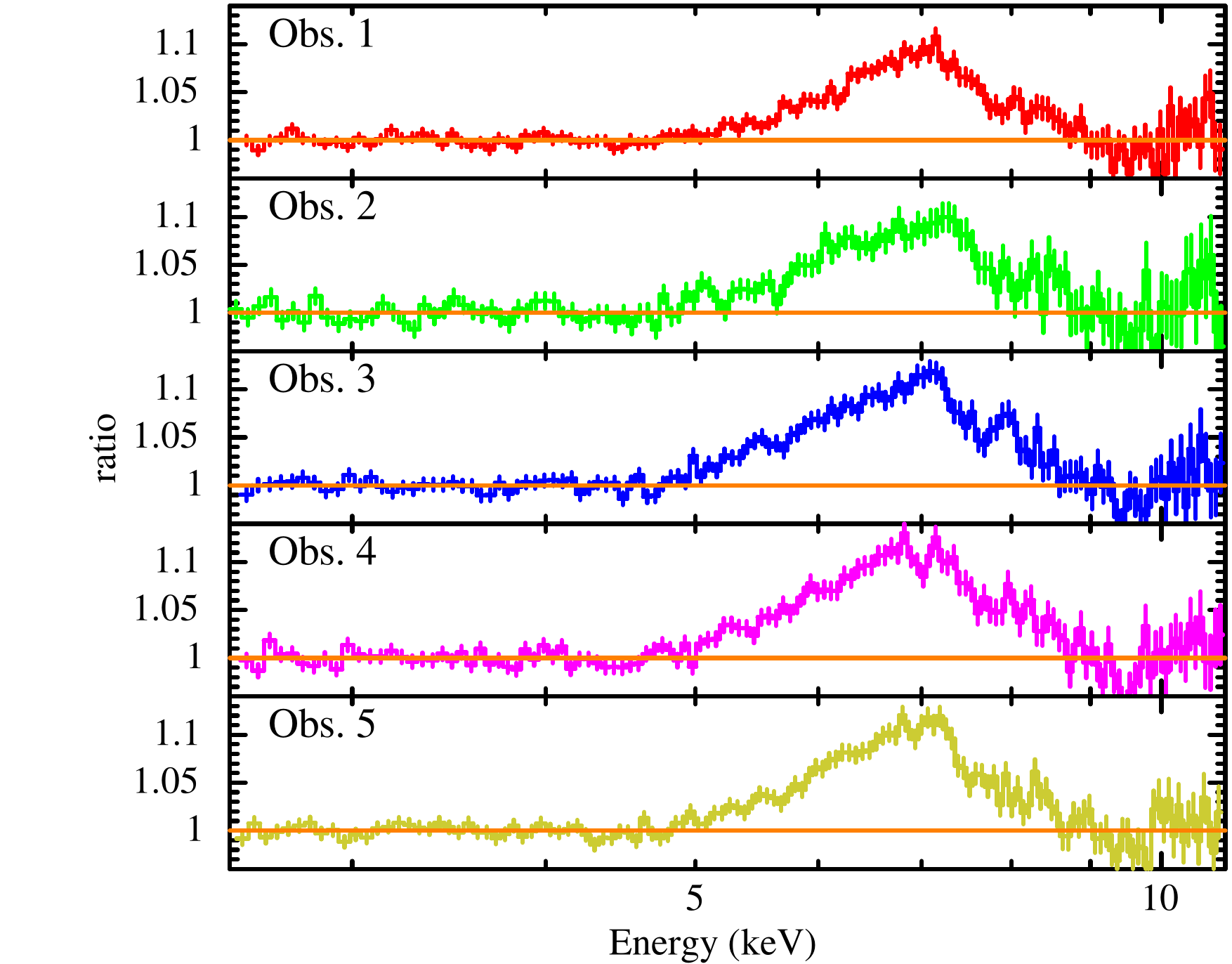}
\caption{Data-to-model ratio plots for the five {\xmm}/\textsc{PN} spectra of {\sub} fitted with the model {\sc tbabs*(bbodyrad+nthcomp)} over the energy ranges of 2.5--5~keV and 8.5--11~keV.}
\label{fig:ratio}
\end{figure}

We then fitted the spectra with the same components, but only in the energy ranges of 2.5--5 and 8.5--11~keV; we show the data-to-model ratio of individual observation in Fig.~\ref{fig:ratio}. A strong broad asymmetric emission feature appears at around 5--9~keV in each spectrum in this plot.

During this fit, we also found that: (1) the seed photon temperature of the {\sc nthcomp} component, $kT\rm_{seed}$, in all the spectra, except for Obs.~1, is consistent with 0; we therefore linked this parameter across the spectra of Obs.~2 to Obs.~4 and got an upper limit at 0.4~keV. However, in order to use a value that was consistent with the one used in the models that we applied in the following sections, we fixed this parameter at $kT\rm_{seed}=0.05$~keV. This improved the constraints on other parameters without extra effect on the fit. The electron temperature of the {\sc nthcomp} component, $kT\rm_{e}$, pegged at its upper limit, 1000~keV, in the spectra of Obs.~1--4 and we thus fixed $kT\rm_{e}$ at 300~keV in these observations to be consistent with the value that is required by the other models (see details in the following sections). Both the seed photon temperature in Obs.~1 and the electron temperature in Obs.~5 of the {\sc nthcomp} component were free to vary.
If we change {\sc nthcomp} to {\sc cutoffpl} to describe the hard part of the spectrum, we obtain a worse fit in this case; the $\chi^2$ increased from $\chi^2=469.2$ with $\nu=374$ to $\chi^2=516$ for $\nu=375$.

\begin{figure}
\centering  
\hspace{-0.5cm}
\subfloat{\includegraphics[width=3.35cm,angle=270]{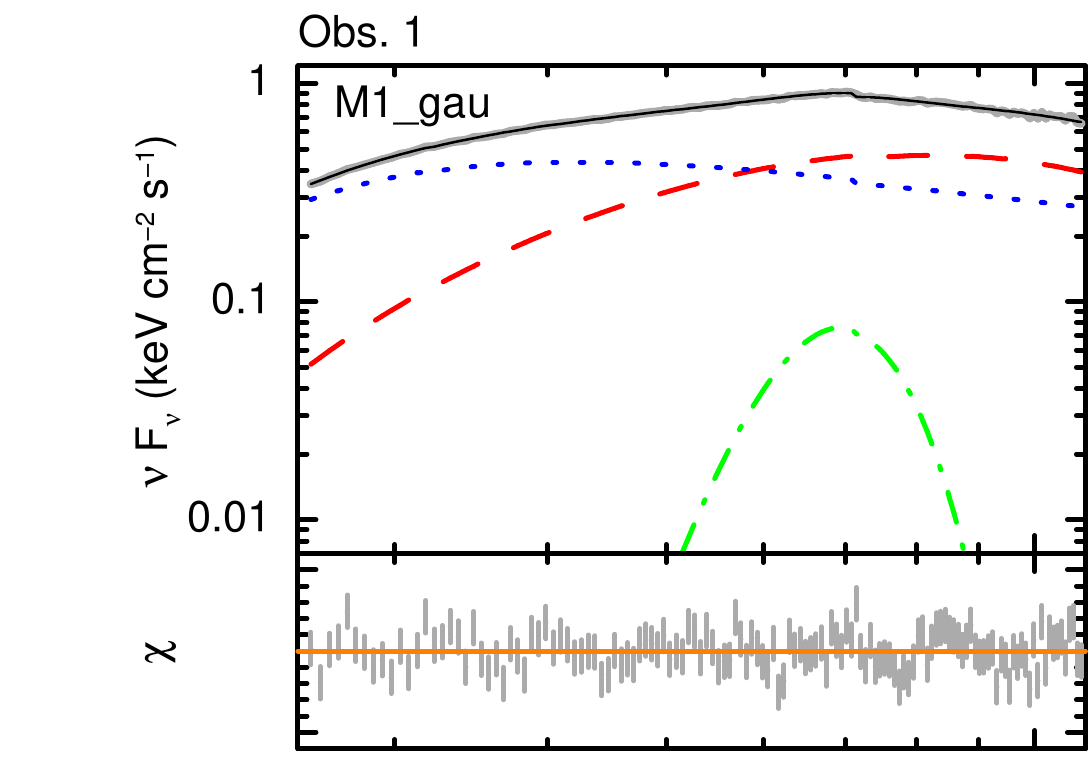}}
\hspace{-1.15cm}
\qquad
\subfloat{\includegraphics[width=3.35cm,angle=270]{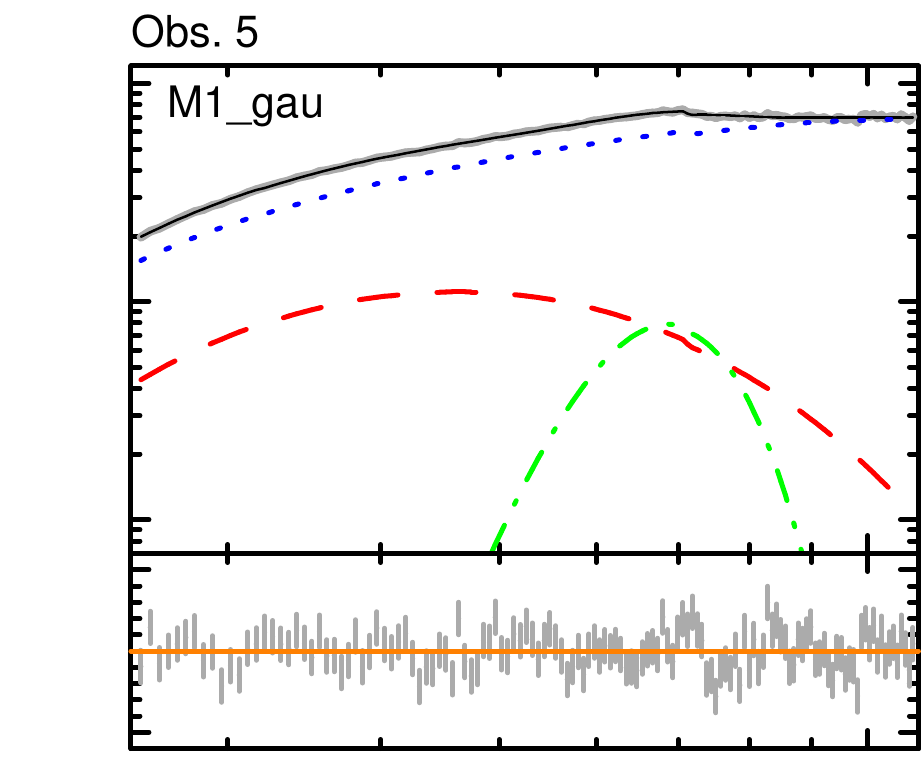}}

\vspace{-0.05cm}
\hspace{-0.5cm}
\subfloat{\includegraphics[width=3.1cm,angle=270]{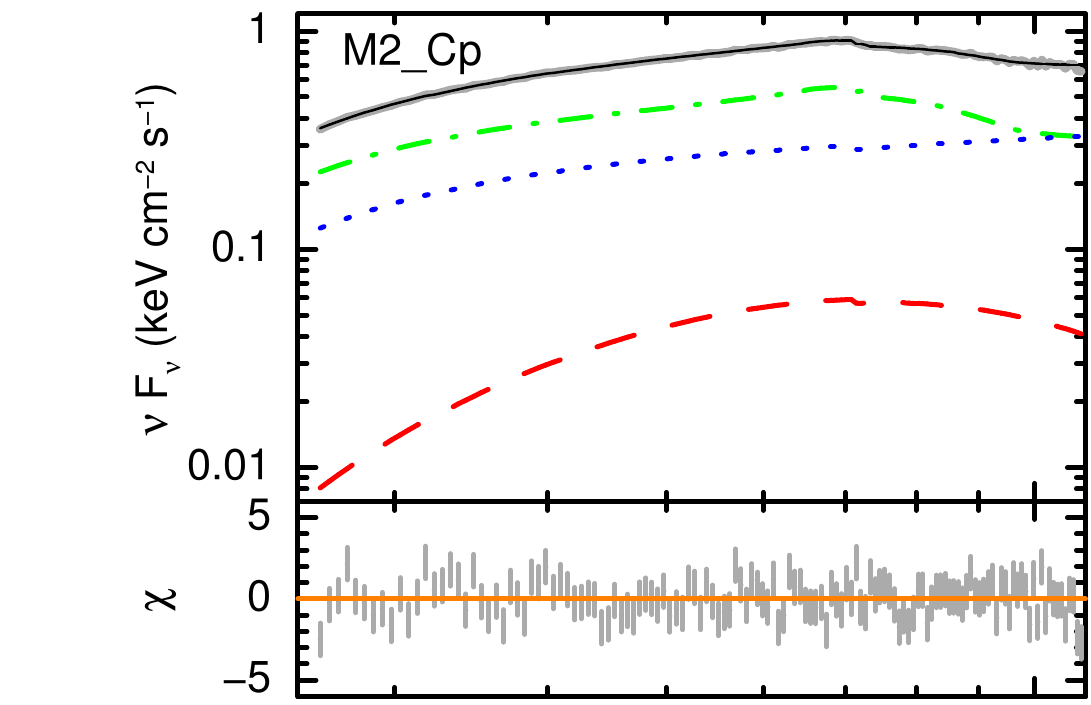}}
\hspace{-1.15cm}
\qquad
\subfloat{\includegraphics[width=3.1cm,angle=270]{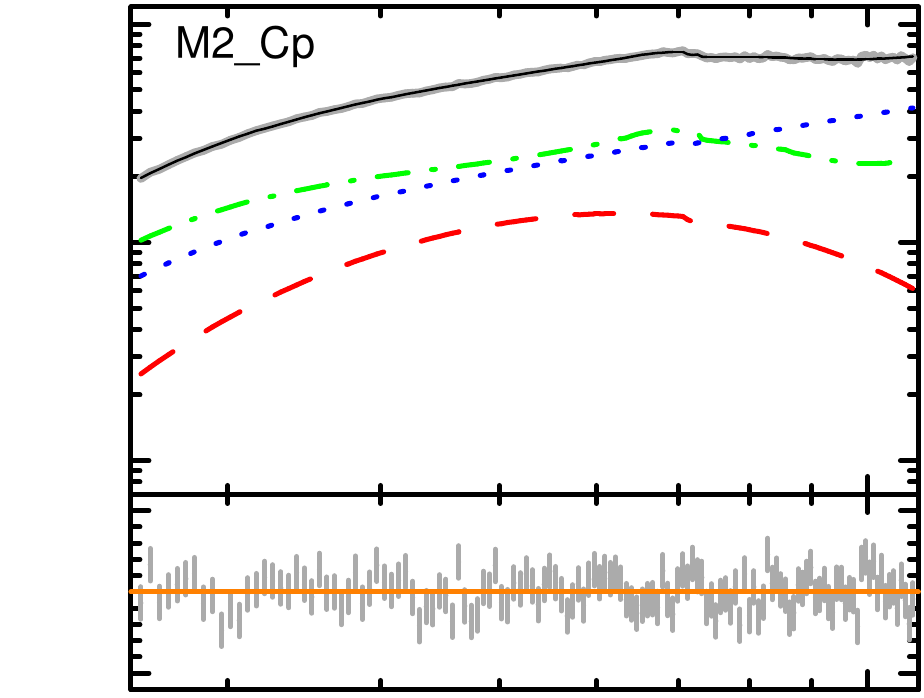}}

\vspace{-0.05cm}
\hspace{-0.5cm}
\subfloat{\includegraphics[width=3.1cm,angle=270]{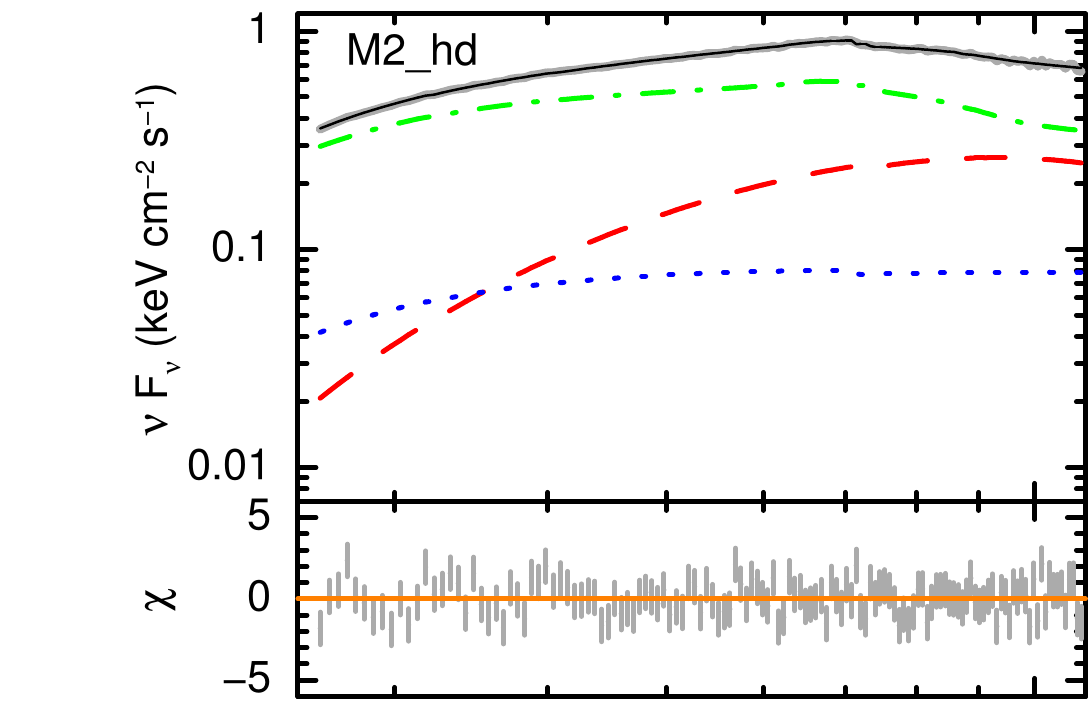}}
\hspace{-1.15cm}
\qquad
\subfloat{\includegraphics[width=3.1cm,angle=270]{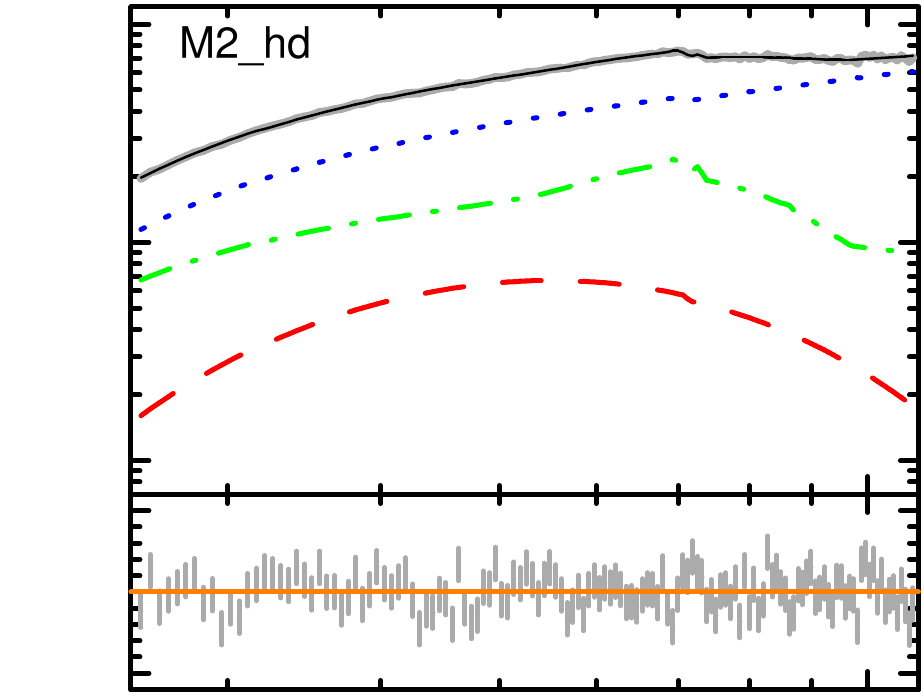}}

\vspace{-0.05cm}
\hspace{-0.5cm}
\subfloat{\includegraphics[width=3.1cm,angle=270]{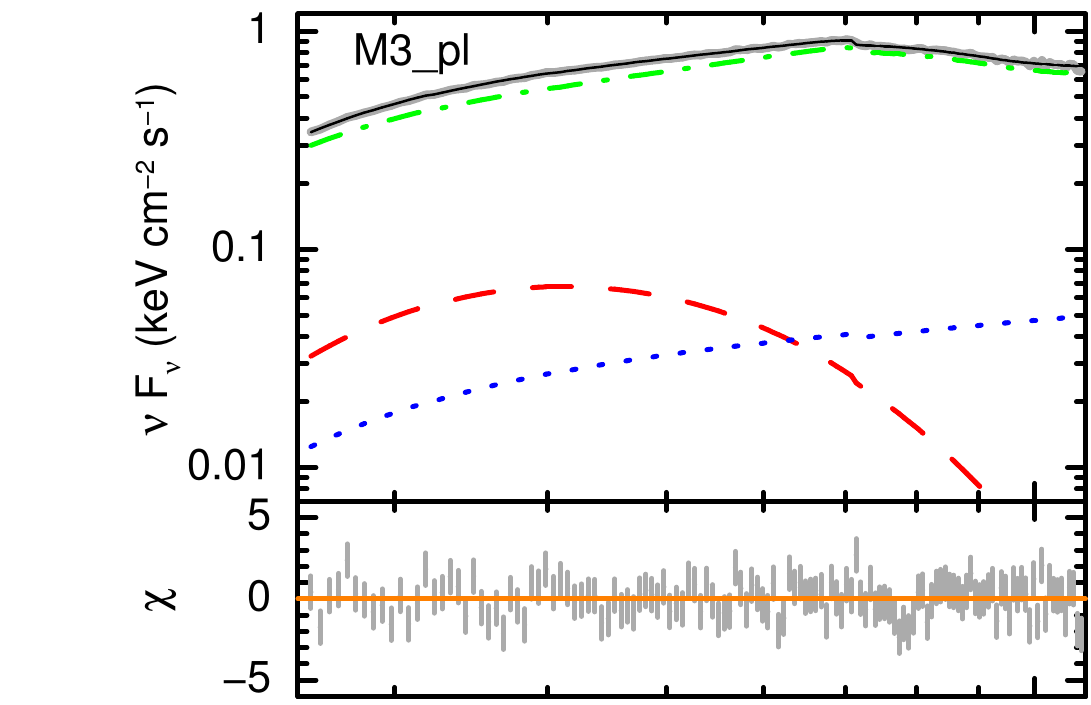}}
\hspace{-1.15cm}
\qquad
\subfloat{\includegraphics[width=3.1cm,angle=270]{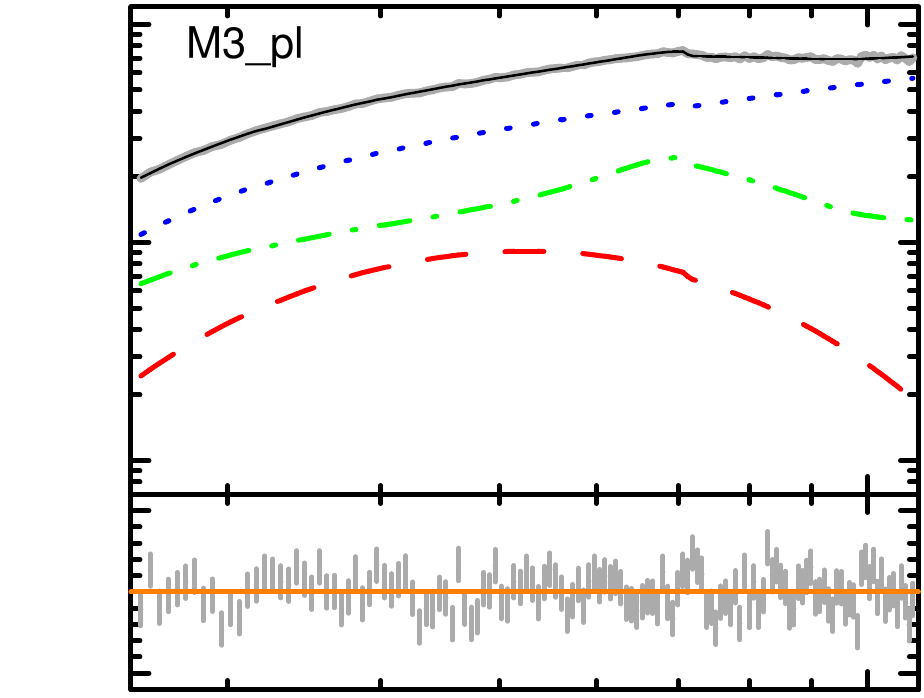}}
\qquad

\vspace{-0.05cm}
\hspace{-0.5cm}
\subfloat{\includegraphics[width=3.1cm,angle=270]{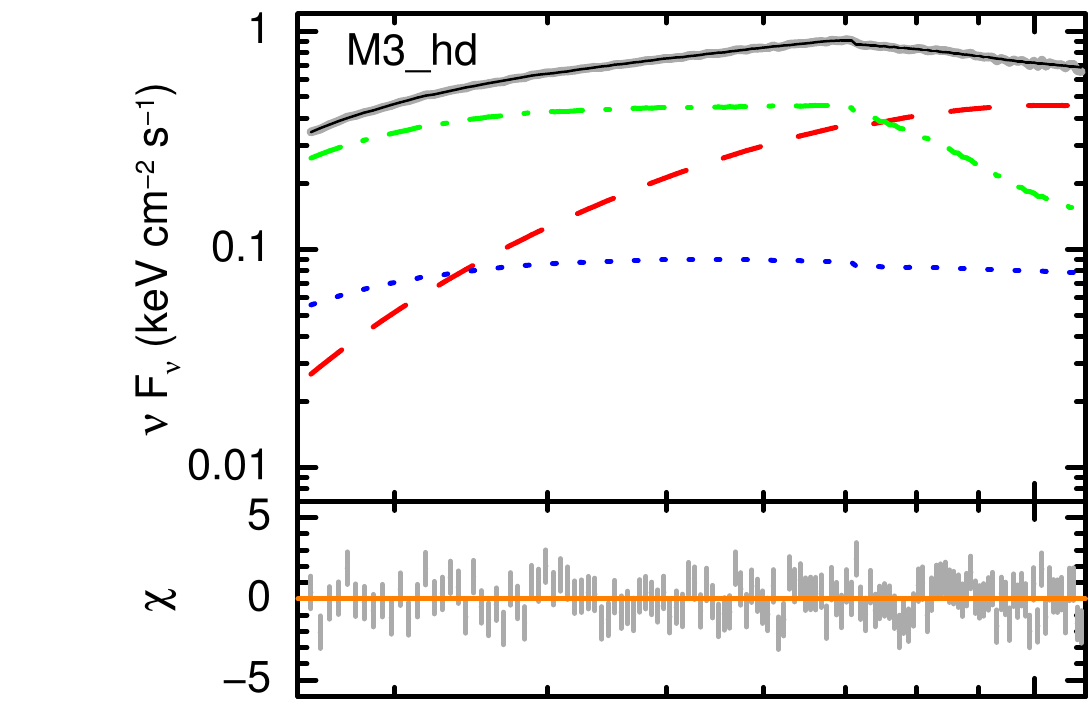}}
\hspace{-1.15cm}
\qquad
\subfloat{\includegraphics[width=3.1cm,angle=270]{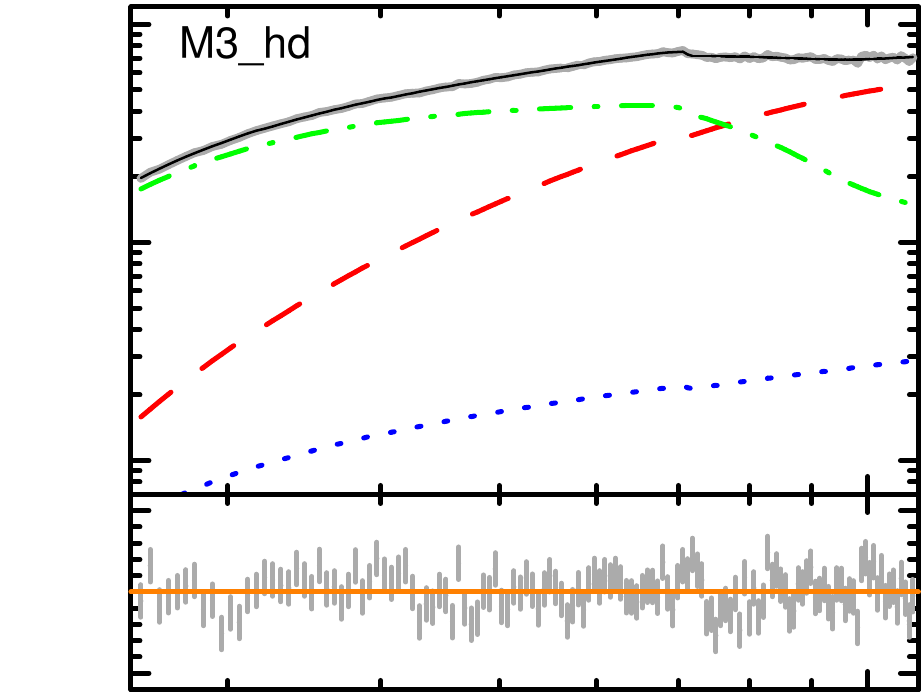}}

\vspace{-0.05cm}
\hspace{-0.5cm}
\subfloat{\includegraphics[width=3.1cm,angle=270]{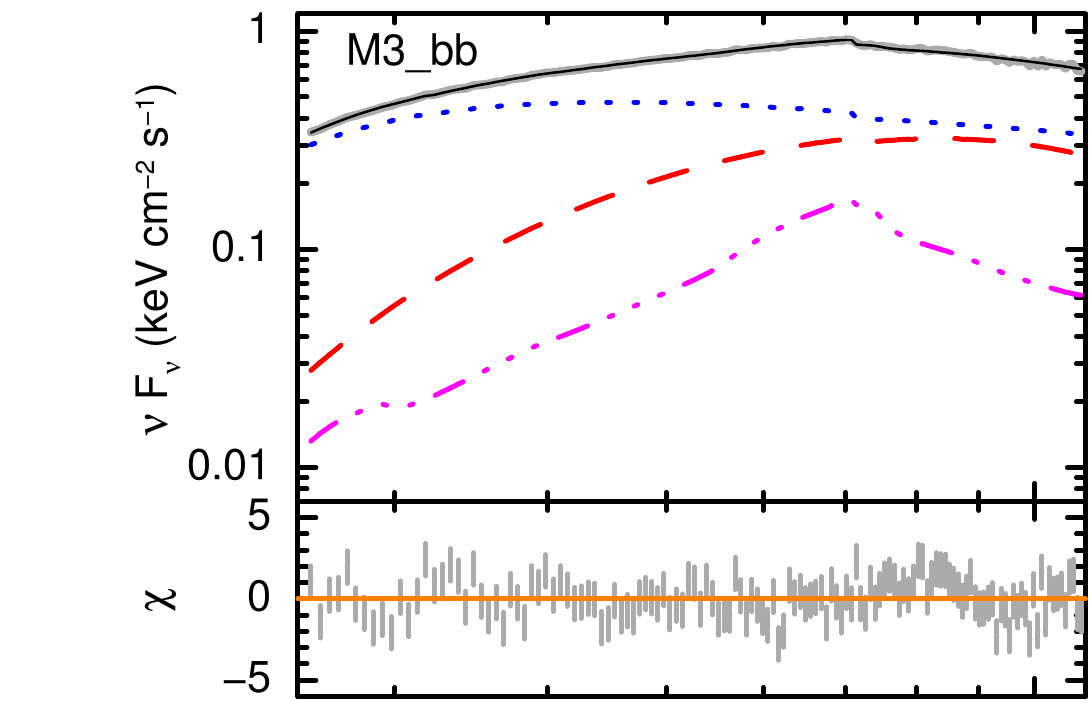}}
\hspace{-1.15cm}
\qquad
\subfloat{\includegraphics[width=3.1cm,angle=270]{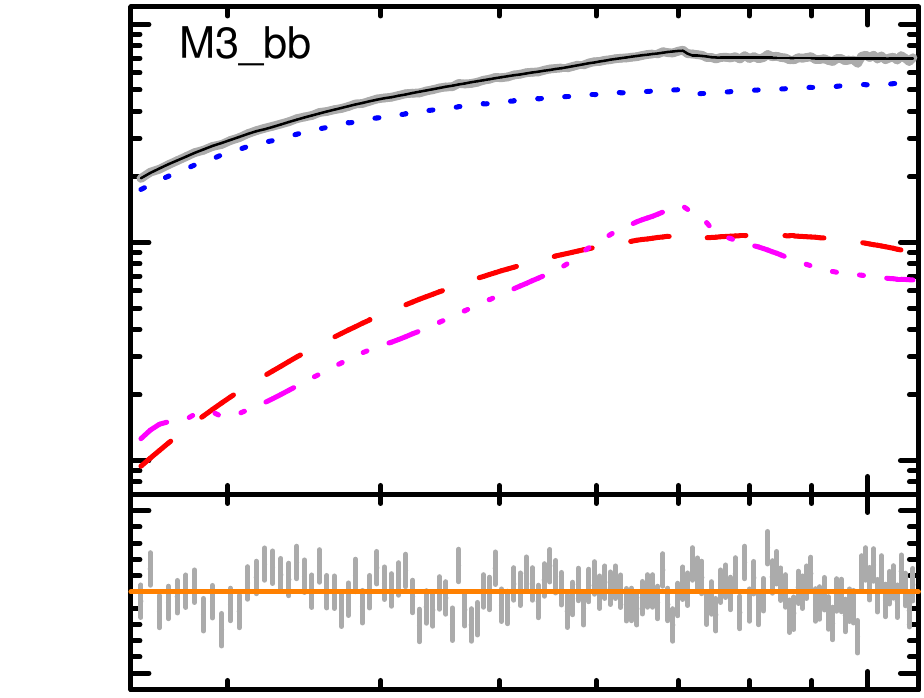}}

\vspace{-0.05cm}
\hspace{-0.5cm}
\subfloat{\includegraphics[width=3.55cm,angle=270]{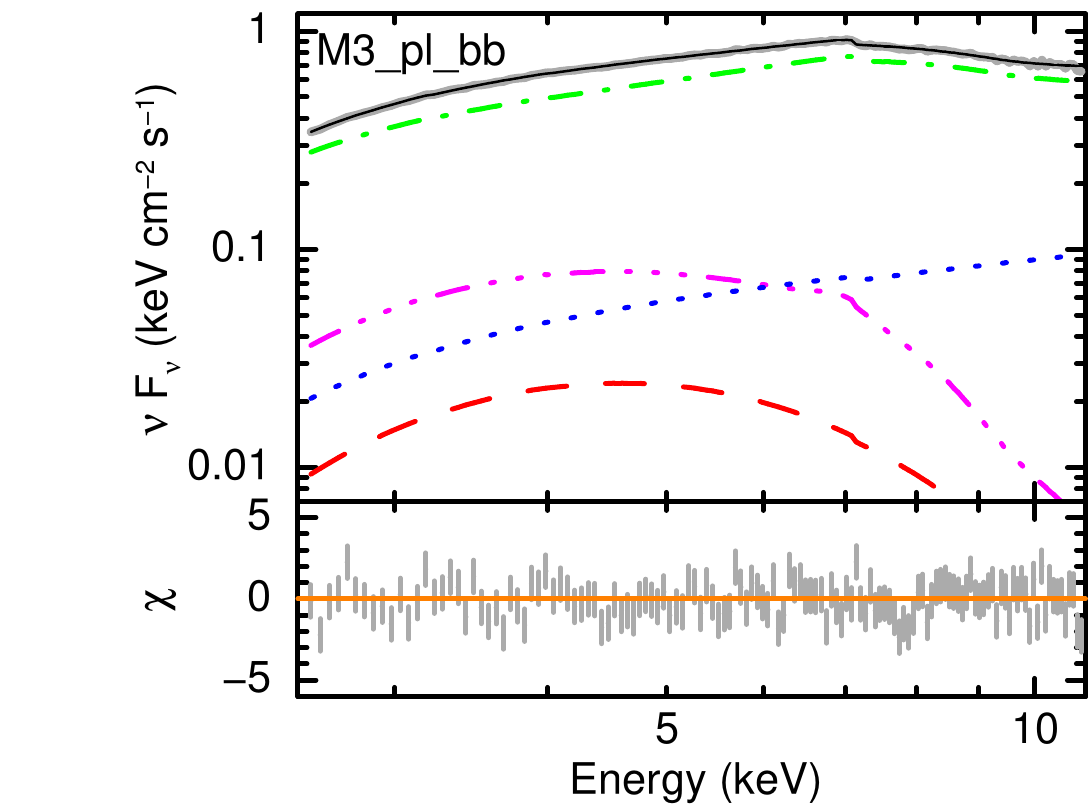}}
\hspace{-1.15cm}
\qquad
\subfloat{\includegraphics[width=3.55cm,angle=270]{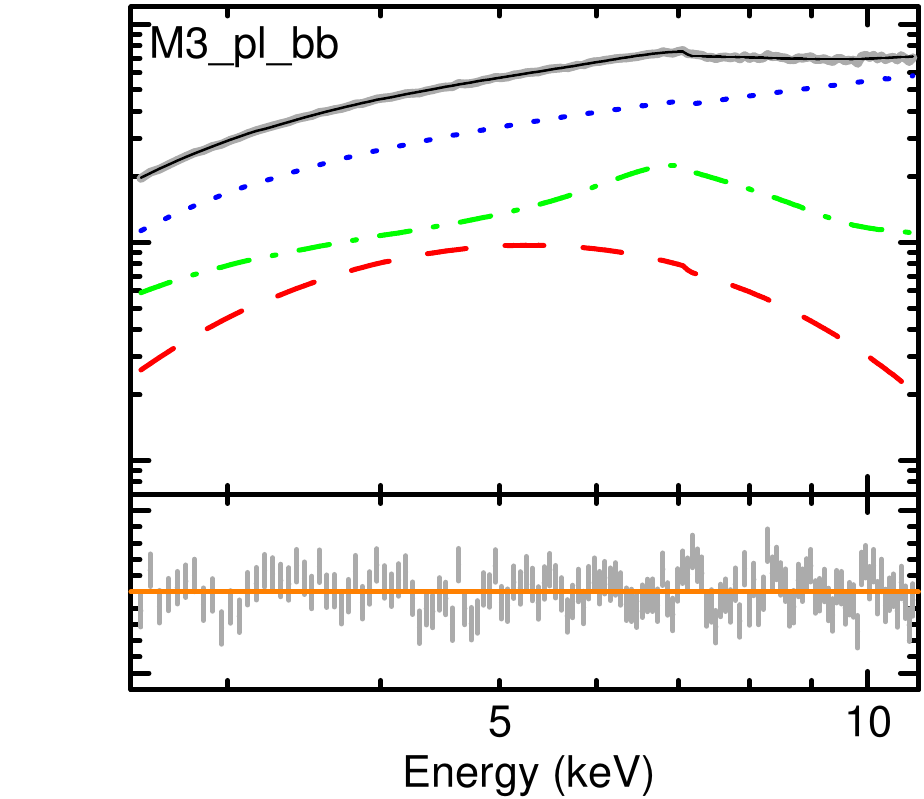}}

\caption{The unfolded spectra and components of different models for Obs.~1 and 5. The residuals in units of sigmas of the fits are shown in the lower panels. The red dashed, blue dotted, green/magenta dotted-dashed and black solid curves show the single blackbody, the Comptonised, the Gaussian/reflection and the entire model, respectively.}
\label{fig:uf_chi}
\end{figure}

Using this continuum model, we fitted the broad emission feature with a simple {\sc gaussian} component. In this case we fitted the data over 2.5--11~keV the full energy range. We call this model M1\_gau; the fit yields $\chi^2=770.1$ for $\nu=638$ and with a null hypothesis probability of $2.4\times10^{-4}$. The seed temperature $kT\rm_{seed}$, in {\sc nthcomp} is $0.70\pm0.05$~keV in Obs.~1 and the electron temperature $kT\rm_{e}$, in {\sc nthcomp}, is $3.4\pm0.1$~keV in Obs.~5. 
As an example, we show the individual components and model residuals in terms of sigmas for Obs.~1 and 5 in the upper panels Fig.~\ref{fig:uf_chi}, since Obs.~1 and 5 represent respectively the softest and hardest spectra of the source in our samples. 

\begin{figure} 
\centering  
\hspace{-1cm}
\resizebox{1\columnwidth}{!}{\rotatebox{0}{\includegraphics[clip]{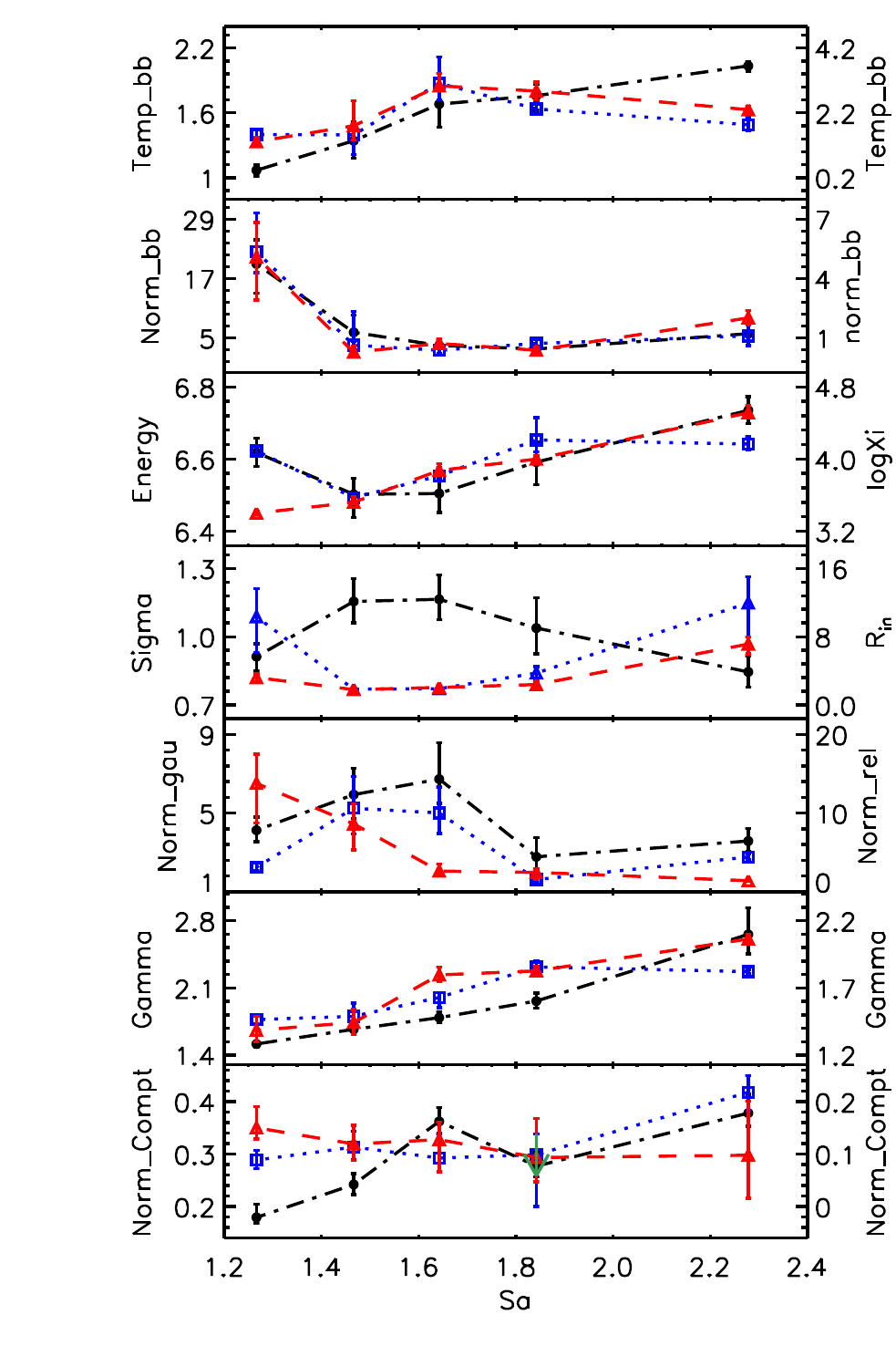}}}
\vspace{-4mm}
\caption{Changes of the best-fitting parameters of the {\xmm} spectra of {\sub} as a function of $S_a$ for M1\_gau (black dot/dashed-dotted line, left y-axis), M2\_Cp (blue square/dotted line, right y-axis) and M2\_hd (red triangle/dashed line, right y-axis). 
From the top to the bottom panels the parameters are the blackbody temperature (keV) and the normalisation ($R_{\rm km}^{2}/D_{10}^{2}$, where $R\rm_{km}$ is the source radius in km and $D_{10}$ is the distance to the source in units of 10~kpc), the line energy (keV)/the disc ionisation ($\rm erg~cm~s^{-1}$), the line width (keV)/the disc inner radius ($R\rm_{ISCO}$), the {\sc gaussian} normalisation ($10^{-3}$)/the {\sc relxillCp}/{\sc relxillD} normalisation ($10^{-3}$), the photon index and the {\sc nthcomp} normalisation, respectively. The green arrow indicates the 95\% confidence upper limit of the {\sc relxillCp} normalisation of Obs.~2.}
\label{fig:para_all}
\end{figure}

\begin{figure} 
\centering  
\hspace{-1cm}
\resizebox{0.85\columnwidth}{!}{\rotatebox{0}{\includegraphics[clip]{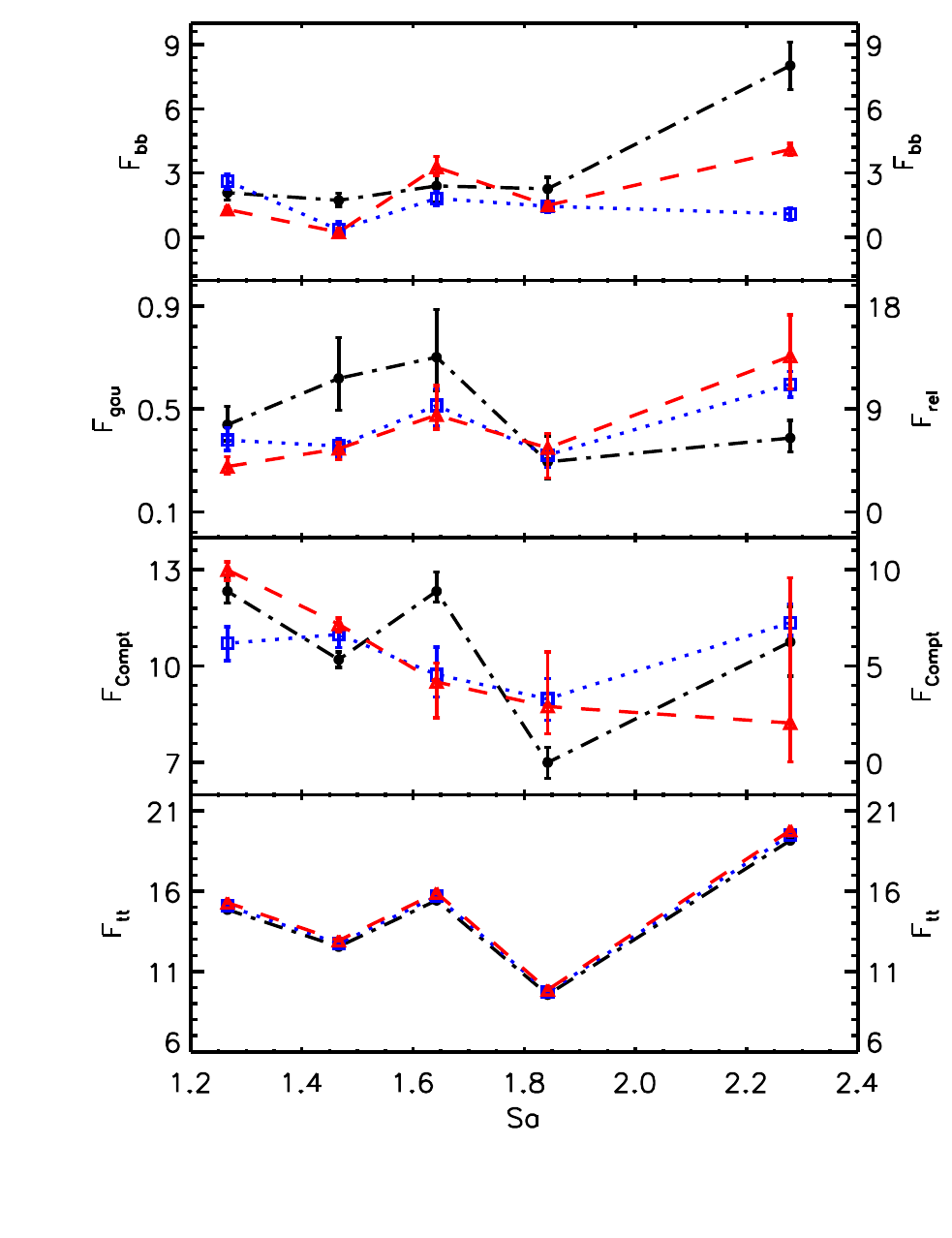}}}
\vspace{-0.9cm}
\caption{The unabsorbed flux of each component for the fit of the spectra of {\sub} with the models M1\_gau (black dot/dashed-dotted line, left y-axis), M2\_Cp (blue square/dotted line, right y-axis) and M2\_hd (red triangle/dashed line, right y-axis).
From the top to the bottom panels, $F\rm_{bb}$, $F\rm_{gau}$/$F\rm_{rel}$, $F\rm_{Compt}$ and $F\rm_{tt}$ represent, respectively, the unabsorbed fluxes of the components {\sc bbodyrad}, {\sc gaussian}/{\sc relxillCp}/{\sc relxillD}, {\sc nthcomp}/{\sc cutoffpl} and the entire model in the 2.5--11~keV range in units of $\rm 10^{-10}~erg~cm^{-2}s^{-1}$. Errors are quoted at the 1~$\sigma$ confidence level.}
\label{fig:flux}
\end{figure}

We report the best-fitting parameters and the individual flux of each component of M1\_gau in Table~\ref{tab:gau} and show the evolution of the parameters and flux of each individual component as a function of $S_{a}$ in Figs.~\ref{fig:para_all} and \ref{fig:flux}, respectively.
The photon index $\Gamma$, in {\sc nthcomp}, and the blackbody temperature $kT\rm_{bb}$, increased monotonically with $S_a$, consistent with the softening of the spectrum as the source evolved in the CD.
The centroid energy of the {\sc gaussian} component decreased first and then increased, while the width of the line changed in the opposite way.
The fluxes of the {\sc nthcomp}, $F_{\rm Compt}$, and the {\sc gaussian}, $F_{\rm gau}$, components change in correlation with each other except in Obs.~2, indicating that the corona was probably the main illuminating source of the reflection component, here represented by the {\sc gaussian} line.
Even though the flux of the soft component, $F_{\rm bb}$, in Obs.~1 in the 2.5--11~keV energy band is almost four times higher than that in other observations, the hard {\sc nthcomp} component dominates the emission during the entire evolution; the total flux, $F_{\rm tt}$, peaks in Obs.~1 and does not appear to change in a simple manner with the source state.

In order to test if adding the RGS data to the fits can help constraining the column density, we fitted the RGS spectra in the energy range between 1 and 2~keV, together with \textsc{PN} spectra in the energy range 2.5--11~keV. 
For the two RGS and one \textsc{PN} spectra of the same observation, we tied all parameters to each other, with two multiplicative factors, one for each RGS instrument, left free to vary; for the same instrument, among different observations, this multiplicative factors were linked. 
The best-fitting value of the column density, $N\rm_{H}=4.1\pm0.03\times10^{22}~cm^{-2}$, obtained for the joint fit of the RGS and \textsc{PN} spectra, is similar with the value, $N\rm_{H}=4.5\pm0.1\times10^{22}~cm^{-2}$, that we derived from the fit to the \textsc{PN} spectra only. 
In the end, we found that adding the RGS data did not improve the value of $N_{\rm H}$ significantly, and therefore we continued using the \textsc{PN} spectra only.

In previous studies the best-fitting hydrogen column density along the line of sight was $2.4-2.6\times10^{22} \rm~cm^{-2}$ (e.g., \citealt{Dai2006,Egron2011}). However, since in these papers they used different cross-sections and solar-abundance tables from ours, it is no surprising that the column density in our case is not consistent with theirs. \cite{Mondal2017} and \cite{Sleator2016} analyzed {\Nu} and {\sw} data of {\sub} using the same cross-sections and solar-abundance tables as ours to calculate the column density, and found the column density as $N_{\rm H}\sim 3.9-4.5 \times 10^{22}~\rm cm^{-2}$.

\begin{table*}
\caption{\label{tab:gau} Best-fitting parameters of M1\_gau, {\sc tbabs*(bbodyrad+gaussian+nthcomp)}}
\renewcommand{\arraystretch}{1.3}
\footnotesize
\centering
\begin{tabular}{clcccccc}
\hline \hline
\multicolumn{2}{c}{Components }& Obs.~1 &  Obs.~2 & Obs.~3 & Obs.~4 & Obs.~5 \\

\hline
{\sc tbabs} &$N\rm_{H}~(10^{22}~cm^{-2})$ &$4.45_{-0.11}^{+0.05l}$&..&..&..&..\\
\hline
{\sc bbodyrad} &$T\rm_{bb}$~(keV)& $2.03\pm0.04$ &$1.76\pm0.09$ &$1.68\pm0.11$ &$1.34\pm0.06$ &$1.07\pm0.03$ \\
&$N\rm_{bb}~(km^{2}/100~kpc^2$)& $5.84_{-0.79}^{+0.21}$ &$2.77_{-0.16}^{+0.08}$ &$3.47\pm0.38$ &$6.12_{-0.85}^{+0.71}$ &$20.0_{-2.6}^{+3.1}$ \\
&flux&  8.0$\pm1.1$& 2.3$\pm0.6$ &2.4$\pm0.8$  & 1.7$\pm0.4$ & 2.1$\pm0.4$ \\
{\sc gaussian} &$E\rm_{gau}~(keV)$ &$6.73_{-0.05}^{+0.01}$ &$6.59_{-0.07}^{+0.01}$ &$6.50\pm0.03$ &$6.50\pm0.03$ &$6.62\pm0.04$ \\
&$\sigma\rm~(keV)$&$0.85_{-0.03}^{+0.09}$ &$1.04\pm0.09$ &$1.16\pm0.05$ &$1.15_{-0.08}^{+0.01}$ &$0.91\pm0.07$ \\
&$N\rm_{gau}~(10^{-3})$&$3.6_{-0.3}^{+0.8}$ &$2.8\pm0.2$ &$6.7\pm0.7$ &$5.9_{-1.0}^{+0.2}$ &$4.1\pm0.6$ \\
&flux& $0.4\pm0.1$ &$0.3\pm0.1$& $0.7\pm0.2$ & $0.6\pm0.2$ &$0.4\pm0.1$\\
{\sc nthcomp} &$\Gamma$&$2.66\pm0.12$ &$1.96\pm0.05$ &$1.79\pm0.05$ &$1.67\pm0.01$ &$1.52\pm0.02$ \\
&$kT\rm_{e}~(keV)$&$300^{f}$&..&..&..&$3.4\pm0.2$\\
&$kT\rm_{bb}~(keV)$&$0.70\pm0.05$&$0.05\pm0.02^l$&..&..&..\\
&$N\rm_{nth}$&$0.38\pm0.02$ &$0.28\pm0.02$ &$0.36\pm0.01$ &$0.24\pm0.01$ &$0.18\pm0.01$ \\
&flux& 10.8$\pm1.1$ & 7.0$\pm0.5$  &12.3$\pm0.5$  & 10.2$\pm0.4$  &12.3$\pm0.4$  \\
\hline
total flux && 19.2$\pm0.1$ & 9.6$\pm0.1$  & 15.4$\pm0.1$  & 12.6$\pm0.1$  &14.9$\pm0.1$  \\
$\chi^2/\nu$&&\multicolumn{5}{c}{$770.1/638$}\\
null hypothesis probability && \multicolumn{5}{c}{$2.4\times10^{-4}$}\\
\hline
\end{tabular}
\begin{flushleft}
{\bf Note:} In this and the following tables, the symbol $l$ indicates that the parameters are linked across the observations, $f$ means that the parameter is fixed during the fit, $p$ denotes that the parameter pegged at its limit and $u$ stands for 95\% confidence upper limit. All the unabsorbed fluxes are in units of $10^{-10}\rm erg~cm^{-2}s^{-1}$ in the 2.5--11~keV range. Errors are quoted at the $1\sigma$ confidence level. 
\end{flushleft}
\end{table*}

\subsubsection{\bf Relativistic reflection model}
Since the plots in Fig.~\ref{fig:ratio} suggest that the broad profile of the emission line at 7~keV is not symmetric and it has been argued in the past that this may be due to Doppler and relativistic effects, we fitted the spectra with the self-consistent reflection model {\sc relxillCp}\footnote{http://www.sternwarte.uni-erlangen.de/$\sim$dauser/research/relxill/} v0.5b \citep{Dauser2014,Garc2014}. This component includes the thermal Comptonisation model {\sc nthcomp} as the illuminating continuum. To limit the number of the free parameters we set the inner and outer emissivity indices of this component to be the same time, we set both of them to be the same, $q\rm_{in}$~=~$q\rm_{out}$, and let $q\rm_{in}$ free to vary across observations.
Following \cite{Braje2000} and assuming a 1.4~$M_{\odot}$ NS, we adopted
a dimensionless spin parameter $a_{*} = 0.47/P\rm_{ms}$, where $P\rm_{ms}$ is the spin period in ms. Since the spin frequency of {\sub} is 363~Hz \citep{Strohmayer1996}, we fixed $a_{*} = 0.17$.

Our result of the fit of the model M1\_gau in the previous section showed that the best-fitting values of the electron temperature were much larger than the upper bound of the \textsc{PN} energy range in all observations except for Obs.~5. To make a fair comparison between models, as the high energy rollover is fixed at 300~keV by default in {\sc relxillD} (and also in {\sc reflionx}-based models, which we will apply in the following sections), we fixed the electron temperature in {\sc relxillCp} at 300~keV in Obs.~1--4.
Once we got a good fit with the model {\sc tbabs*(bbodyrad+relxillCp)}, we froze the parameter reflection fraction, refl\_frac, at its negative value to force this component to only account for the reflected part, we added the direct {\sc nthcomp} component to the model and linked the common parameters, the photon index and the electron temperature, of both the components {\sc nthcomp} and {\sc relxillCp}. This procedure allows us to get the fluxes of the individual components separately. We followed the same procedure when we used other {\sc relxill}-based models in this paper. 

The overall model, hereafter M2\_Cp, became {\sc tbabs*(bbodyrad+relxillCp+nthcomp)}, which yields $\chi^2/\nu=685.4/631$. Compared with the fit with M1\_gau, the $\chi^2$ of this fit decreased by $\Delta\chi^2=84.7$ for 7 d.o.f. less. 
The emissivity index was not well constrained and was marginally consistent within errors in all the observations. We therefore linked this parameter across observations to improve the constraint on other parameters, which yields $\chi^2/\nu=694.6/635$ and with a null hypothesis probability of 0.05 (see the unfolded spectra and models in Fig.~\ref{fig:uf_chi}).

\begin{figure} 
\centering  
\hspace{-0.5cm}
\subfloat{\includegraphics[width=3.2cm,angle=270]{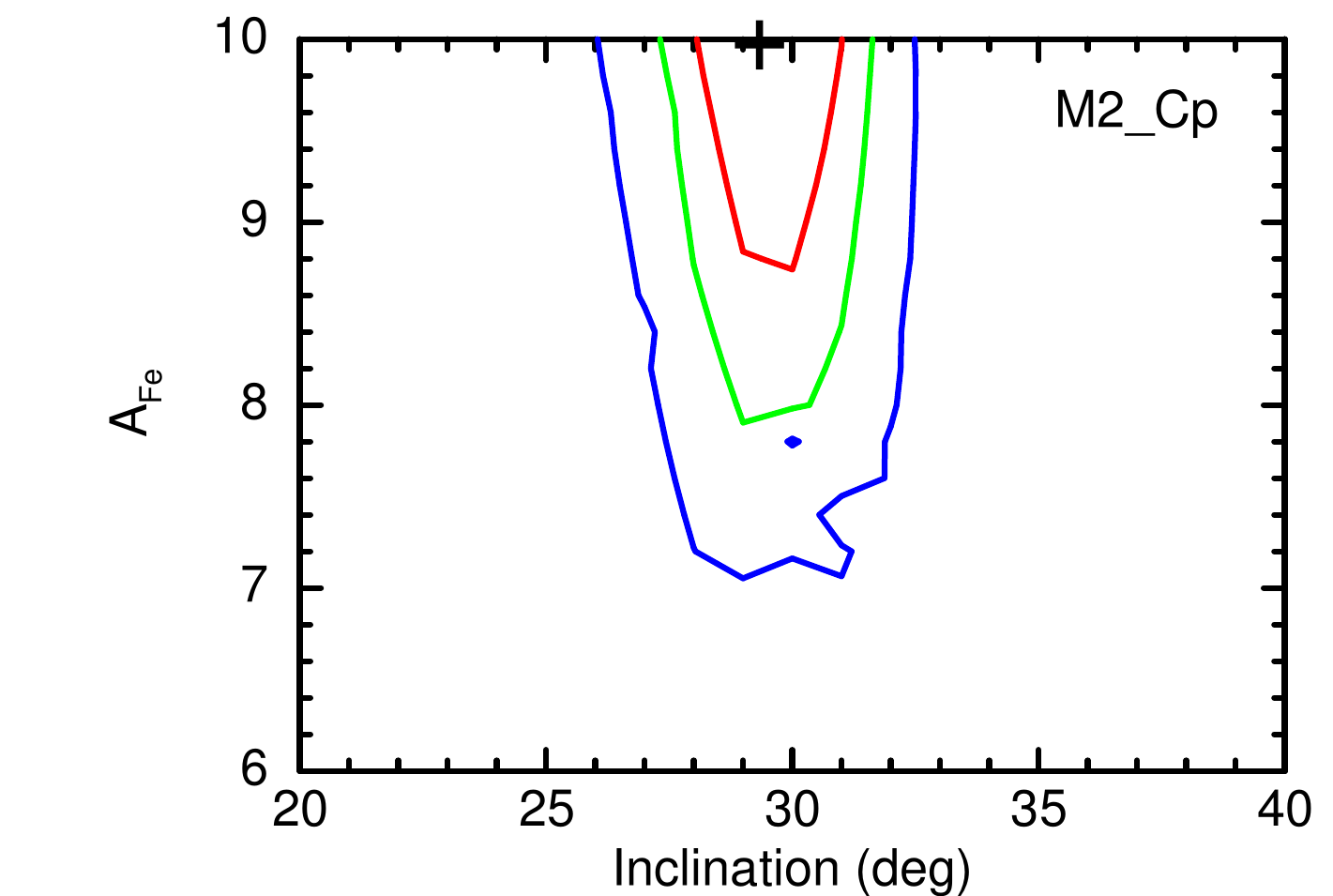}}
\hspace{-1.25cm}
\qquad
\subfloat{\includegraphics[width=3.2cm,angle=270]{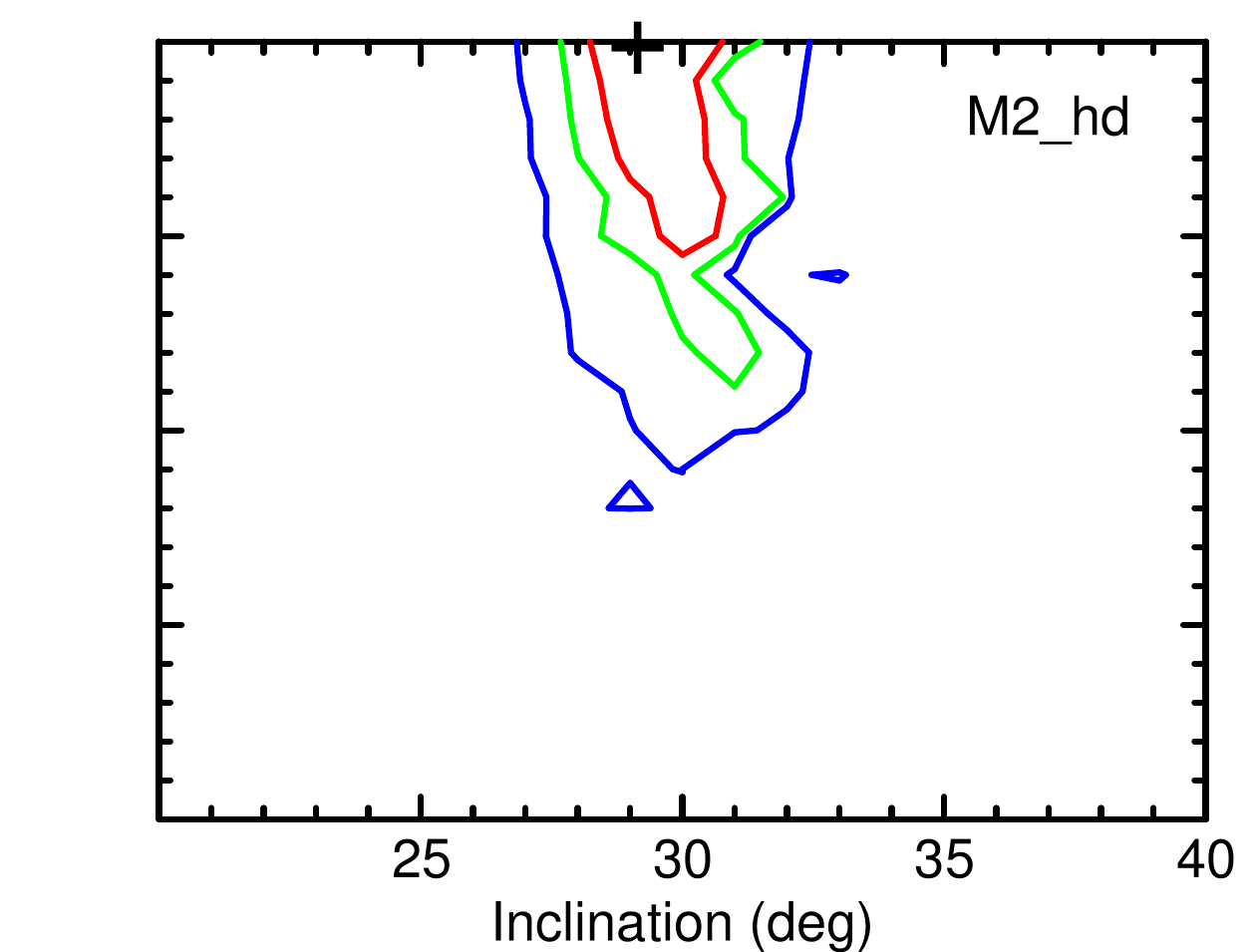}}
\caption{Contour plots for iron abundance vs. inclination at the 68\% (red), 90\% (green) and 99\% (blue) confidence levels for models M2\_Cp (left panel) and M2\_hd (right panel). The best-fitting values of both parameters are marked with a cross.}
\label{fig:chi}
\end{figure}

We show the relevant parameters of this model in Table~\ref{tab:relcp} and plot some of the parameters of each component as a function of $S_a$ in Fig.~\ref{fig:para_all}. In Obs.~1, 2 and 3, the spectrum is dominated by the reflection component, whereas in Obs.~4 and 5 the fluxes of the reflection and the Comptonised components are comparable.
There are, however, two issues with this fit: (1) the best-fitting value of the iron abundance is very high, $A\rm_{Fe}=10$ times solar abundance, which is the upper bound of this parameter (see the contour plot for the iron abundance vs. the inclination in Fig.~\ref{fig:chi}); (2) some of the best-fitting parameters in this model are not consistent with the same parameters in M1\_gau, e.g., both the blackbody temperature and the photon index in M1\_gau monotonously increase with $S_a$ whereas the same parameters in M2\_Cp first increase and then decrease or remain more or less constant. If we forced the iron abundance to be 1, the fit becomes worse and $\chi^2$ increases by $\Delta \chi^2/\Delta\nu=100.9/1$.

In all the Cp-type versions of the {\sc relxill}-based models, the seed temperature is fixed at 0.05~keV by default, which is more than 10 times smaller than the best-fitting value of $kT\rm_{seed}$ that we obtained from M1\_gau in Obs.~1. This discrepancy may partly cause the inconsistent results between M1\_gau and M2\_Cp.

We also tried to fit the data with other types of the {\sc relxill}-based reflection models: the fit with the model {\sc relxill} was almost as good as the one with M2\_Cp, $\chi^2/\nu=697.1/635$; as for the lamppost model, {\sc relxilllpCp}, the fit yielded a similar $\chi^2$, $\chi^2/\nu=696.2/635$. However, as the Compton hump is not covered by {\xmm}, we cannot constrain the key parameter, the height of the corona, in this model well. It is worth noting that the iron abundance, $A_{\rm Fe}=10$ times solar abundance, pegs at its upper limit in the fits with all these reflection models.

\begin{table*}
\caption{\label{tab:relcp} Best-fitting parameters for M2\_Cp, {\sc tbabs*(bbodyrad+relxillCp+nthcomp)}}
\renewcommand{\arraystretch}{1.3}
\footnotesize
\centering
\begin{tabular}{clcccccc}
\hline \hline
\multicolumn{2}{c}{Components }& Obs.~1 &  Obs.~2 & Obs.~3 & Obs.~4 & Obs.~5 \\

\hline
{\sc tbabs} &$N\rm_{H}~(10^{22}~cm^{-2})$ &$4.92\pm0.10^{l}$&..&..&..&..\\
\hline
{\sc bbodyrad} &$T\rm_{bb}$~(keV)& $1.83_{-0.18}^{+0.25}$ &$2.32_{-0.18}^{+0.26}$ &$3.11_{-0.52}^{+0.81}$ &$1.52_{-0.61}^{+0.33}$ &$1.53\pm0.15$ \\
&$N\rm_{bb}~(km^{2}/100~kpc^2$)& $1.1\pm0.6$ &$0.7\pm0.2$ &$0.4\pm1.8$ &$0.6\pm0.13$ &$5.3_{-1.1}^{+2.0}$ \\
&flux& $1.1\pm0.3$ & $1.5\pm0.3$ & $1.8\pm0.3$ & $0.4\pm0.3$ & $2.6\pm0.4$ \\
{\sc relxillCp} &$q\rm_{in}$&$3.9\pm0.6^{l}$ &..&..&..&..\\
& $A\rm_{Fe}$ &$10_{-0.7}^{+0p}$&..&..&..&..\\
& $i$~($\degree$) &$29.6\pm1.0^{l}$&..&..&..&..\\
& $a_{*}$ &$0.17^{f}$&..&..&..&..\\
&$kT\rm_{e}~(keV)$&$300^{f}$&..&..&..&$8.9\pm0.3$\\
&$\Gamma$&$1.82\pm0.02$ &$1.86\pm0.05$ &$1.63\pm0.06$ &$1.49\pm0.09$ &$1.46\pm0.03$ \\
&$R\rm_{in}$~$(R\rm_{ISCO})$&$12.0_{-6.1}^{+3.1}$ &$3.7\pm0.9$ &$1.9\pm0.2$ &$1.9\pm0.2$ &$10.3_{-4.2}^{+3.3}$ \\
&$\log~\xi~$$(\rm erg~cm~s^{-1})$&$4.17\pm0.07$ &$4.21_{-0.13}^{+0.25}$ &$3.81\pm0.04$ &$3.57\pm0.05$ &$4.09\pm0.05$ \\
&$^\natural$refl\_frac& $0.9\pm0.5$ &$1.2_{-0.7}^{+5.5}$&$0.5_{-0.2}^{+0.7}$&$0.5\pm0.2$&$0.6\pm0.3$\\
&$N\rm_{relcp}~(10^{-3})$&$4.4\pm0.4$ &$1.6\pm0.5$ &$10.0_{-2.6}^{+3.3}$ &$10.6_{-3.3}^{+4.1}$ &$3.1\pm0.5$ \\
&flux& $11.2\pm1.1$ & $5.0_{-1.1}^{+1.9}$ & $9.3\pm1.6$ & $5.8\pm0.8$ & $6.3\pm1.0$ \\
{\sc nthcomp} &$N\rm_{nth}$&$0.2\pm0.04$ &$<0.1^{u}$ &$0.09\pm0.05$ &$0.1\pm0.03$ &$0.09\pm0.02$ \\
&flux& $7.3\pm0.9$ & $3.3\pm1.1$ & $4.6\pm1.3$ & $6.6\pm0.6$ & $6.2\pm0.9$ \\
\hline
total flux&& $19.5\pm0.1$ & $9.7\pm0.04$ & $15.7\pm0.1$ & $12.7\pm0.1$ & $15.1\pm0.1$ \\
$\chi^2/\nu$&&\multicolumn{5}{c}{$694.6/635$}\\
null hypothesis probability && \multicolumn{5}{c}{$5.0\times10^{-2}$}\\
\hline
\end{tabular}
\begin{flushleft}
{\bf Note:} All the symbols and units are the same as in Table~\ref{tab:gau}. $^\natural$The parameter, refl\_frac, has been frozen at its negative value to force the component {\sc relxillCp} to only account for the reflection part.
\end{flushleft}
\end{table*}

\subsubsection{\bf Reflection model with high density}
A high iron abundance using these reflection models has been reported in previous works and, in some cases, the authors argued that this was the result of the low density of the accretion disc, $n\rm_{e} = 10^{15}~\rm cm^{-3}$ used in the calculation of the models (e.g., \citealt{Garca2016, Tomsick2018}).
To test the possible effect of the density on the disc iron abundance,
we fitted the spectra with the extended reflection model {\sc relxillD} \citep{Garca2016} that allows the electron density parameter to vary between $n\rm_{e} = 10^{15}$ and $10^{19}\rm~cm^{-3}$, which we call M2\_hd.
We replaced {\sc relxillCp} by {\sc relxillD}, {\sc nthcomp} by {\sc cutoffpl} with the cut-off energy, $E\rm_{cut}$, fixed at 300~keV since this is required by the {\sc relxillD} component, and applied the same fixed parameters as for M2\_Cp. 
The electron density, $\log(n_{\rm e})$, was linked to be the same in Obs.~1--5.

The best-fitting parameters and individual unabsorbed flux are given in Table~\ref{tab:reld} and are added as red triangles to Figs.~\ref{fig:para_all} and ~\ref{fig:flux}. 
Compared to the fit with M2\_Cp, the new fit slightly improved, such that $\chi^2$ decreased by $\Delta\chi^2=7.1$ for the same $\nu$ and with a null hypothesis probability of 0.073. The iron abundance still pegged at 10, with a high density of $\log(n_{\rm e})$ up to 19; the evolution of the photon index in M2\_hd is similar to that in M1\_gau. The right panel in Fig.~\ref{fig:chi} shows the contour plot for the iron abundance vs. the inclination for M2\_hd.
If we force the iron abundance to be 1, the fit becomes worse, with $\Delta \chi^2=131.9$ for $\Delta\nu=1$.
Similar to M2\_Cp, the flux of M2\_hd in Obs.~1--3 is dominated by the reflection component, whereas in Obs.~4 and 5 it is dominated by the Comptonised component.

\begin{table*}
\caption{\label{tab:reld} Best-fitting parameters for M2\_hd, {\sc tbabs*(bbodyrad+relxillD+cutoffpl)}}
\renewcommand{\arraystretch}{1.3}
\footnotesize
\centering
\begin{tabular}{clcccccc}
\hline \hline
\multicolumn{2}{c}{Components }& Obs.~1 &  Obs.~2 & Obs.~3 & Obs.~4 & Obs.~5 \\

\hline
{\sc tbabs} &$N\rm_{H}~(10^{22}~cm^{-2})$ &$5.28\pm0.04^{l}$&..&..&..&..\\
\hline
{\sc bbodyrad} &$T\rm_{bb}$~(keV)& $2.30\pm0.07$ &$2.87\pm0.27$ &$3.03_{-0.12}^{+0.38}$ &$1.80_{-0.44}^{+0.78}$ &$1.31_{-0.08}^{+0.14}$ \\
&$N\rm_{bb}~(km^{2}/100~kpc^2$)& $2.0\pm0.4$ &$0.4\pm0.2$ &$0.7\pm0.2$ &$0.3\pm0.3$ &$5.1_{-2.2}^{+1.7}$ \\
&flux& $4.1\pm0.3$ &$1.5\pm0.2$&$3.3\pm0.5$&$0.2\pm0.2$&$1.3\pm0.1$\\
{\sc relxillD} &$q\rm_{in}$ &$3.5\pm0.2^{l}$&..&..&..&..\\
& $A\rm_{Fe}$ &$10_{-0.4}^{+0p}$&..&..&..&..\\
& $\log(n_{\rm e})~(cm^{-3})$ &$19_{-0.1}^{+0l}$&..&..&..&..\\
& $i$~($\degree$) &$29.4\pm0.9^l$&..&..&..&..\\
& $a_{*}$ &$0.17^{f}$&..&..&..&..\\
&$E\rm_{cut}~(keV)$&$300^{f}$&..&..&..&..\\
&$\Gamma$&$2.06\pm0.03$ &$1.83\pm0.04$ &$1.80\pm0.06$ &$1.44\pm0.10$ &$1.39\pm0.10$ \\
&$R\rm_{in}$~$(R\rm_{ISCO})$&$7.1\pm1.1$ &$2.4\pm0.3$ &$2.1\pm0.2$ &$1.8\pm0.3$ &$3.2\pm0.4$ \\
&$\log~\xi~$$(\rm erg~cm~s^{-1})$&$4.51_{-0.04}^{+ 0.09}$ &$4.00\pm0.04$ &$3.87\pm0.06$ &$3.52_{-0.01}^{+0.06}$ &$3.40\pm0.02$ \\
&$N\rm_{relD}~(10^{-3})$&$1.4\pm0.1$ &$2.5\pm0.7$ &$2.6_{-0.3}^{+0.9}$ &$8.6_{-3.3}^{+2.6}$ &$13.8_{-5.0}^{+3.7}$ \\
&$^\natural$refl\_frac&$2.6\pm0.2$&$1.0\pm0.3$&$1.0_{-0.2}^{+0.7}$&$0.3\pm0.02$&$0.2\pm0.01$\\
&flux&  13.6$_{-2.9}^{+3.7}$&  $5.6_{-2.7}^{+1.3}$& $8.5_{-1.2}^{+2.6}$ & 5.5$\pm0.8$  & 4.0$\pm0.8$ \\
{\sc cutoffpl} &$N\rm_{pl}$&$0.1\pm0.09$ &$0.09\pm0.06$ &$0.13\pm0.05$ &$0.12\pm0.04$ &$0.15\pm0.04$ \\
&flux& 2.1$_{-0p}^{+7.5}$  &2.9$_{-1.4}^{+2.8}$  &4.2$_{-1.9}^{+1.0}$  & 7.2$\pm0.4$ &10.0$\pm0.5$  \\
\hline
total flux&& 19.8$\pm0.1$  &9.9$\pm0.02$  & 15.9$\pm0.04$ & 12.9$\pm0.03$  &15.3$\pm0.04$  \\
$\chi^2/\nu$&&\multicolumn{5}{c}{$687.5/635$}\\
null hypothesis probability && \multicolumn{5}{c}{$7.3\times10^{-2}$}\\
\hline
\end{tabular}
\begin{flushleft}
{\bf Note:} All the symbols and units are the same as in Table~\ref{tab:gau}. $^\natural$The parameter, refl\_frac, has been frozen at its negative value to force the component {\sc relxillD} to only account for the reflection part.
\end{flushleft}
\end{table*}

\subsubsection{\bf Alternative reflection model}
To check the robustness of the values we obtained from the fits with the {\sc relxill}-based models, and especially to explore the issue of the supersolar iron abundance, we also fitted the data with the model {\sc reflionx} \citep{Ross2005} that characterizes the emergent reflection spectrum arising from an illuminating power-law spectrum, including a high-energy exponential cutoff with e-folding energy fixed at 300~keV; we convolved this component with the relativistic convolution model {\sc kerrconv} \citep{Brenneman2006}. The model that we fitted in XSPEC was {\sc tbabs*(bbodyrad+kerrconv*reflionx+cutoffpl)}, which we call M3\_pl. 

As in the previous fits, in {\sc reflionx} we tied the inner and outer emissivity indices, $q\rm_{in}$~=~$q\rm_{out}$, and set the spin parameter $a_{*}=0.17$. The value of $q\rm_{in}$ in M3\_pl was consistent with being the same in all observations, so we linked this parameter across all the observations. Additionally, we linked the photon index and the cut-off energy in {\sc reflionx} to those in {\sc cutoffpl}.
The final fit is worse than M2\_Cp, $\chi^2/\nu=727.3/637$ and with a null hypothesis probability of $6.8\times10^{-3}$; the parameters are listed in Table~A1 in the appendix. The best-fitting iron abundance is $5.8_{-0.04}^{+0.7}$ and the inclination angle of the accretion disc with respect to the line of sight is $24.6\pm1.2$.
The average blackbody temperature for M3\_pl is smaller than for M2\_Cp, but the trends of the photon index and the inner radius are similar to those for M2\_Cp. 
The reflection flux for M3\_pl in Obs.~1-3 is larger than in the rest of the observations; the reflection and Comptonised fluxes in Obs.~4 are almost equal, and the Comptonised flux in Obs.~5 is dominant.

We also applied the high electron density version of this model, {\sc reflionx\_hd} (M3\_hd, \citealt{Tomsick2018}), in which the density in the reflector can go up to $10^{22}~\rm cm^{-3}$; the iron abundance is fixed at the solar abundance. The overall fit is worse than M3\_pl, $\chi^2/\nu=763.5/637$ and with a null hypothesis probability of $3.7\times10^{-4}$; the corresponding best-fitting parameters are shown in Table~A2. Compared to M2\_hd, the evolution of the inner radius of both models is analogous. The spectrum fitted with M3\_hd is dominated by the reflection component all the time.

Since {\sub} is a NS, we tried to fit the reflection spectrum with another version of {\sc reflionx}, {\sc reflionx\_bb} (\citealt{Ludlam2017}), in which the illuminating source is the blackbody component. In this model, {\sc tbabs*(bbodyrad+kerrconv*reflionx\_bb+cutoffpl)}, which we call M3\_bb, we linked the blackbody temperature, $kT$ in {\sc reflionx\_bb}, to the blackbody temperature, $kT_{\rm bb}$, in {\sc bbodyrad} in all observations. 
The fit is worse than M3\_pl, $\chi^2/\nu=753.3/637$ and with a null hypothesis probability of $3.9\times10^{-5}$ (see Table~A3), suggesting that the illuminating source in {\sub} cannot only be the blackbody component. However, different from the above results, the iron abundance derived from M3\_bb is $0.78_{-0.09}^{+0.01}$ and the inclination angle is $52.9_{-0.5}^{+1.6}$. Two other differences are that the column density and the overall blackbody temperature in M3\_bb are higher than those in M3\_pl. The spectrum for this model is dominated by the Comptonised component in all observations.

To further identify whether the illuminating source is the corona or the NS surface/the boundary, we combined the Comptonised {\sc reflionx} and the blackbody {\sc reflionx\_bb} versions together in a model, M3\_pl\_bb. We assumed that the iron abundance and the ionization of the disc in both components are the same. In Obs.~5 the normalization of the {\sc reflionx\_bb} component is negligible, as well as the normalization of the {\sc reflionx} component in Obs.~2. We show the parameters in Table~A4 and the fit yields $\chi^2/\nu=730.2/634$ and with a null hypothesis probability of  $3.2\times10^{-3}$, similar with M3\_pl, although the inner radius in Obs.~1 is very large, $R_{\rm in}=57~R_{\rm ISCO}$, and the iron abundance is consistent with the one in M3\_pl. The flux for Obs.~1 and 3 is dominated by the {\sc reflionx} component; the flux for Obs.~2 and 5 is dominated by the Comptonised component. In Obs.~4 the fluxes of the {\sc reflionx} and the Comptonised components are almost the same. 
Except in Obs.~2, the {\sc reflionx} flux is always larger than that of the {\sc reflionx\_bb} component.


\subsection{Tests with {\Nu} data}
As previously mentioned in Section~\ref{intro}, {\sub} was also observed with {\Nu}. \cite{Mondal2017} analysed two {\Nu} observations (ObsIDs: 80001012002 and 80001012004) plus two simultaneous {\sw}/XRT observations (ObsIDs: 00080185001 and 00080185002) and found an iron abundance $A_{\rm Fe}=~$2--5 times solar. To test if the discrepancy in the iron abundance derived from our and their models is due to the lack of coverage of the high energy range (above $\sim \rm 11~keV$) in our data, we re-analysed {\Nu} observation 80001012002 and the simultaneous XRT observation 00080185001 in which the source was in the hard state.
We used M2\_Cp to jointly fit the {\Nu} observation in the energy ranges 3.5--50.0/3.5--11.0~keV and the XRT observation in the energy range 1.0--7.5~keV. Type I bursts were detected and removed from the {\Nu} spectra. Although M2\_Cp and M2\_hd fit the {\xmm} data equally well, a cut-off energy is required by the {\Nu} spectra \citep{Mondal2017} and the cut-off energy is frozen at 300~keV as default in {\sc relxillD}, therefore here we chose M2\_Cp to do this test.

In Table~\ref{tab:nusw} we show the best-fitting results when the photons above 11~keV are either included or excluded in the {\Nu} spectra. The results show that most of the parameters are marginally consistent with each other no matter whether the hard photons are included in the spectra or not; as expected, the parameters that are affected the most are the photon index, $\Gamma$, and the electron temperature, $kT_{\rm e}$, from the {\sc nthcomp} component. 
On the other hand, both the reflection and the Comptonised components are less well constrained when we exclude the hard photons. Another significant difference is that the iron abundance, $A_{\rm Fe}$, increases from $\sim2$ times solar when we include the {\Nu} data above 11 keV, to $\sim8$ times solar when we fit the {\Nu} spectra only in the 3.5--11~keV range.

\begin{table}
\caption{\label{tab:nusw} Best-fitting parameters for the {\Nu} and {\sw} data with M2\_Cp}
\renewcommand{\arraystretch}{1.3}
\footnotesize
\centering
\begin{tabular}{clcccccc}
\hline \hline
\multicolumn{2}{c}{Components/in the energy ranges of}& 3.5--50~keV &  3.5--11~keV \\
\hline
{\sc const}&$\rm {FPMA}$&$1^f$&$1^f$\\
&$\rm FPMB$&$1.044\pm0.002$&$1.042\pm0.002$\\
&$\rm XRT$&$1.037\pm0.011$&$1.048\pm0.011$\\
\hline
{\sc tbabs} &$N\rm_{H}~(10^{22}~cm^{-2})$ &$4.9\pm0.1$&$4.6\pm0.2$\\
\hline
{\sc bbodyrad} &$T\rm_{bb}$~(keV)& $2.12\pm0.07$ &$2.32\pm0.04$ \\
&$N\rm_{bb}~(km^{2}/100~kpc^2$)& $1.9\pm0.4$ &$5.0\pm0.5$  \\
{\sc relxillCp} &$q\rm_{in}$ &$2.0_{-0.7}^{+10p}$&$3.4_{-0.8}^{+2.6}$\\
& $A\rm_{Fe}$ &$1.8\pm0.8$&$5.1_{-1.7}^{+3.2}$\\
& $i$~($\degree$) &$39.2_{-14.6}^{+10.0}$&$26.8_{-5.5}^{+3.2}$\\
& $a_{*}$ &$0.17^{f}$&$0.17^{f}$\\
&$kT\rm_{e}~(keV)$&$11.0\pm0.5$&$4.9_{-1.7}^{+3.2}$\\
&$\Gamma$&$1.92\pm0.03$ &$1.60\pm0.07$ \\
&$R\rm_{in}$~$(R\rm_{ISCO})$&$3.8_{-3.8p}^{+29.2}$ &$2.1\pm0.9$ \\
&$\log~\xi~$ $(\rm erg~cm~s^{-1})$&$3.94_{-0.16}^{+ 0.20}$ &$3.89\pm0.12$ \\
&$N\rm_{rel}~(10^{-3})$&$5.0\pm1.3$ &$<6.9^{u}$\\
\hline
{\sc nthcomp} &$N\rm_{nth}$&$0.4\pm0.1$ &$<0.06^{u}$  \\
\hline
&$\chi^2/\nu$&$2807.1/2497$&$1115.9/1007$\\
\hline
\end{tabular}
\begin{flushleft}
{\bf Note:} The energy range of the {\sw}/XRT data used here is always between 1.0 and 7.5~keV; only the energy range of the {\Nu} data changes. All the symbols and units are the same as in Table~\ref{tab:gau}.
\end{flushleft}
\end{table}

\section{Discussion}
We analysed five {\xmm} observations of the NS LMXB source {\sub} obtained in 2011, when the source evolved from the soft to the hard state, to explore how the accretion flow changed between those states. 
A broad emission line at $\sim6.5-6.7$~keV in the spectrum of this source indicates the presence of a reflection component in this system. 
By jointly fitting all the five spectra with several reflection models, we obtained an inclination angle of $25\degree-53\degree$ and an iron abundance of up to 10 times the solar abundance. In what follows, we compare the spectral parameters derived from the fits with different models, identify the kHz QPOs that we observed in the power spectra, and discuss the possible reasons why a supersolar iron abundance appears to be required by the data.

\subsection{Comparisons of all applied the models}
In this paper, we fitted the continuum spectrum of {\sub} with a single temperature blackbody {\sc bbodyrad}, plus a Comptonised component, {\sc nthcomp}/{\sc cutoffpl} depending on the requirement of the model, to account for the soft and hard photons in the spectra, respectively. 
As shown in Fig.~\ref{fig:ratio}, a strong emission feature appears to be present in the 5--9~keV energy range of each spectrum. We used several components to fit this emission: a {\sc gaussian} component in M1\_gau and the reflection components {\sc relxillCp} in M2\_Cp, {\sc relxillD} in M2\_hd, {\sc kerrconv*reflionx} in M3\_pl, {\sc kerrconv*reflionx\_hd} in M3\_hd, {\sc kerrconv*reflionx\_bb} in M3\_bb, {\sc kerrconv*(reflionx+reflionx\_bb)} in M3\_pl\_bb, respectively. 

In M1\_gau, as shown in Table~\ref{tab:gau} and Fig.~\ref{fig:flux}, the line flux is the same in Obs.~1 and 5, which represent, respectively, the softest and the hardest state observations in this work; even though the spectrum was dominated by the hard component all the time, as the total flux decreased, the blackbody flux in Obs.~1 dramatically dropped to one fourth of that in Obs.~5. 

When the emission feature was fitted with the reflection component {\sc relxillCp} in M2\_Cp, 
the spectrum was dominated by the reflection component in Obs.~1--3 and the fluxes of the reflection and the Comptonised components were equally strong in the last two observations. When we took the high-density effect ($>15~\rm cm^{-3}$) into account in the reflection component, the fit with M2\_hd was slightly better than that of M2\_Cp. 
Similar to M2\_Cp in Obs.~1--3, the dominant component in M2\_hd are the reflection component but in Obs.~4--5, the dominant component in M2\_hd turns to be the Comptonised component (see Tables~\ref{tab:relcp} and \ref{tab:reld}). 

Ever though the trends and values of the parameters derived from M2\_Cp and M2\_hd are consistent within errors in Obs.~2--4, these parameters in Obs.~1 and 5 are different.
For instance, the blackbody flux, in units of $10^{-10}\rm erg~cm^{-2}s^{-1}$, of Obs.~1 increased from $1.1\pm0.3$ in M2\_Cp to $4.1\pm0.3$ in M2\_hd and that of Obs.~5 decreased from $2.6\pm0.4$ in M2\_Cp to $1.3\pm0.1$ in M2\_hd. On the contrary, the Comptonised flux, in units of $10^{-10}\rm erg~cm^{-2}s^{-1}$, of Obs.~1 decreased from $7.3\pm0.9$ in M2\_Cp to \textbf{$2.1_{-0p}^{+7.5}$} in M2\_hd and that of Obs.~5 increased from $6.2\pm0.9$ in M2\_Cp to $10.0\pm0.5$ in M2\_hd.

The iron abundance derived from M2\_Cp and M2\_hd both pegs at the upper limit, $A_{\rm Fe}=10$ in solar units. 
When we replaced the self-consistent reflection models {\sc relxillCp} and {\sc relxillD} with the {\sc reflionx}-based components convolved with the relativistic blurring kernel {\sc kerrconv} in M3\_pl, M3\_hd and M3\_bb, the fit became worse, with $\chi^2$ increasing 40--76 for 2 d.o.f. more (see Tables~A1--A3). The iron abundance in M3\_pl and M3\_bb were $5.8_{-0.8}^{+0.04}$ and $0.78_{-0.09}^{+0.01}$ times solar, respectively. When $A_{\rm Fe}$ was forced to be 1 and the density of the disc was allowed to be as high as $10^{22}~\rm cm^{-3}$ in M3\_hd, there was not improvement on the fit. 

The inclination derived from all the models above was around $\sim 30\degree$, except in M3\_bb in which the inclination was $52.9_{-0.5}^{+1.6}$ but the $\chi^2$ of the fit was very large.
Although the version of the combination of the {\sc reflionx} and {\sc reflionx\_bb} components, M3\_pl\_bb, improves the fit compared to the version of the {\sc reflionx\_bb} component alone, this fit does not yield an iron abundance as low as in M3\_bb.

\subsection{Identification of the kHz QPO}
As we mentioned in Section~\ref{sec:qpo}, two single kHz QPOs with a frequency of, respectively, $604 \pm 17$ Hz and $583 \pm 19$~Hz have been detected in the {\rx} data, corresponding to our Obs.~3 with {\xmm}. As reported by \cite{Mendez1999b}, \cite{Salvo2001} and \cite{Straaten2002}, the frequencies of the upper and lower kHz QPOs in {\sub} fall in the range $500-880$~Hz and $380-1160$~Hz, respectively. \cite{Salvo2001} and \cite{Straaten2002} studied the fractional rms amplitude of both kHz QPOs as a function of the QPO frequency. In order to tell whether we have detected the lower or the upper kHz QPOs, we compared the rms amplitude and frequency of our kHz QPOs to the ones in their papers, and found that both our detected QPOs are more likely the upper kHz QPO.

\cite{Mendez1999b} found that, as a function of $S_a$, the frequencies of the upper and lower QPOs follow well-defined separate tracks. If both kHz QPOs that we detected here were the lower kHz QPO, according to Fig.~4 in \cite{Mendez1999b}, the corresponding $S_a$ would be $\sim 1.9$, which is much higher than the one of Obs.~3, $S_a = 1.3$ and, different from what we observe, it would put the source in the transitional intermediate state, close to the vertex of the CD. If the QPOs that we detected were the upper kHz QPO, the corresponding $S_a$ from the same figure would be between 1.3 and 1.4, consistent with our value of $S_a$ for Obs.~3. We therefore conclude that the two detections of kHz QPOs in the {\rx} data that are simultaneous with our {\xmm} Obs.~3 correspond to the upper kHz QPOs in {\sub}.


\subsection{Inner radius uncorrelated with source states}
The evolution of the source on the {\rx} CD and in the {\sw}/BAT light curve in Fig.~\ref{fig:ccd} give an idea of the spectral evolution of {\sub} during the {\xmm} observations presented here, from a relatively soft to the hard state. The evolution of the spectral parameters of M1\_gau support this idea: the blackbody temperature and the photon index of {\sc nthcomp} increase as $S_{a}$ increases, even though the spectra are dominated by the hard component, {\sc nthcomp}, at all times (see Figs~\ref{fig:para_all} and \ref{fig:flux}). The flux of the {\sc gaussian} component followed the same trend as that of the {\sc nthcomp} component, except in Obs.~2.

When we fitted the data with the relativistic reflection models M2\_Cp and M2\_hd, the model parameters follow a similar trend to that of model M1\_gau, except the {\sc relxillD} normalisation in model M2\_hd. 
In the standard truncated accretion disc model, as mass accretion rate increases the inner disc radius moves inwards \citep{Esin1997,Done2007}.
However, Fig.~\ref{fig:para_all} shows that the inner radius derived from both models first decreased from Obs.~1 to 2, it then remains constant in Obs.~2--4 and it finally increased from Obs.~4 to 5. The evolution of the inner radius in Obs.~2--5 supports the truncated disc model above, indicating that the inner radius moves outwards with decreasing mass accretion disc.
However, going from Obs.~1 to 2, with an apparently decreasing mass accretion rate, the inner radius moves inwards.

In the standard accretion disc model, gas pressure dominates when both the accretion rate and the X-ray luminosity ($L_{\rm x}<10^{36}\rm~ergs~s^{-1}$) are low \citep{Shakura1973}. On the contrary, when the luminosity is high, radiation pressure should dominate. \cite{Popham2001} showed that when the luminosity approaches the Eddington limit, the radiation feedback from the NS surface leads to an increase of the inner radius. As the flux of Obs.~1 is the largest one in our samples, this process may result in the inner radius variation that we observe.



\subsection{Iron abundance deduced from {\xmm} and {\Nu} data}
As we showed in Fig.~\ref{fig:chi}, the best-fitting value of the iron abundance in {\sub} from the fits to the {\xmm} data is 10 times solar or higher, which differs from what \cite{Mondal2017} found with {\Nu} and {\sw} data. \cite{Mondal2017} analysed two simultaneous {\Nu} and {\sw} observations carried out in 2013, and inferred that during these two observations the source was in the hard and soft state, respectively.  
Similar to what they did, we assumed that the spin parameter is 0.17, and applied similar models to fit the reflection spectrum: they used {\sc relxill} and we used {\sc relxillCp}; the inclination angle in their and our work are consistent, around $30\degree$, but the iron abundance they obtained is 2--4 times the solar, about half to one fifth of the value that we find.

A high electron density of the accretion disc has been suggested as a potential solution of the supersolar disc iron abundance (e.g. \citealt{Garca2016,Tomsick2018}). \cite{Tomsick2018} explained that a high density produces more soft emission, resulting in a harder power law, which provides a better match to the hard spectrum, as well as an extra soft excess below 1~keV. 
However, compared to the fit with model M2\_Cp, the fit with model M2\_hd that allows for higher density than M2\_Cp, only improved slightly, with the iron abundance pegging at 10 times solar and the density pegging at $10^{19}~\rm cm^{-3}$. Allowing for a higher density of the disc only increased the column density of the interstellar medium and the disc temperature in our fits. 


As the iron abundance derived from model M2\_hd pegged at its upper limit, we tested another model, {\sc reflionx\_hd}, with an electron density that can go up to $10^{22}~\rm cm^{-3}$. Unfortunately, this model, M3\_hd, did not return a good fit and the density still pegged at the upper limit (see Table~\ref{tab:refl_hd}).

Since the iron abundance reported by \cite{Mondal2017} is very different from ours, we did another test with {\Nu} and {\sw} data, as \cite{Mondal2017} used, to see if the lack of the data at energies above 11~keV plays a role in this result. 
As we described in Section~3.3, the iron abundance, $A_{\rm Fe}$, increases from $\sim2$ in solar units when we fit the full {\Nu} data up to 50 keV, to $\sim8$ in solar units when we fit the {\Nu} spectra only in the 3.5--11~keV range. At the same time, the {\sc nthcomp} becomes negligible if we ignore the {\Nu} data above 11~keV. A possible explanation for this result is that in order to produce a similar significant reflection spectrum, more iron is required.  
Even though neither the {\sc relxillCp} nor the {\sc nthcomp} components are well constrained when the hard photons are ignored, we cannot exclude this hypothesis.

\subsection{The possible illuminating source of the reflection component}
Most of the fits show that reflection makes a significant contribution to the entire spectrum. For instance, the reflection component in Obs.~1--3 dominated the total emission in all the models, except in M3\_bb. 
The reflection fraction, refl\_frac, remains constant within errors among the observations, and the reflection flux is independent of the Comptonised flux in the fits with models M2\_Cp and M2\_hd, in both of which we assume that the corona is responsible for the disc reflection. 
We identify two possible explanations for our finding that the changes of the reflection flux and the Comptonised flux are uncorrelated. The first possibility is that both the NS surface/boundary and the corona irradiated the disc and contributed to the reflection spectrum. 
Alternatively, light bending may play a role in the reflection process as well.
As the reflection happens in the vicinity of a compact object, due to the strong gravitational light-bending effect, more of the Comptonised photons would be bent towards the disc, which results in less of the Comptonised photons being observed directly at infinity. \cite{Miniutti2004} identified three different regimes in which the reflection–dominated component (and the iron line) is correlated, anti–correlated or almost independent with respect to the direct continuum, and they concluded that the relation between the reflection and direct component is correlated to the source state and the height of the illuminating source.

\cite{Cackett2010} studied broad iron emission lines in 10 NS-LMXBs and concluded that the boundary layer is the illuminating source irradiating the accretion disc in these systems. In Section~3.2.3 we explored the relative contribution of the corona and the NS surface/boundary layer to the reflection spectrum. 
Comparing the fits with models M3\_pl and M3\_bb, the former gives a better fit, $\Delta \chi^2=25.9$ with the same $\nu$, which suggests that the boundary layer might not be the only contributor to the reflection spectrum in all observations.

Thanks to model M3\_pl\_bb, we can make a direct comparison of the contribution to the reflection spectrum between the corona and the NS surface/boundary layer. 
In Table~A4, we show that the flux of the {\sc reflionx} component is much larger than that of the {\sc reflionx\_bb} component except in Obs.~2. The boundary layer contributed 
4\%--43\% of the total flux to the reflection component in Obs.~1--4, not strong but still required by the data; the contribution of the corona to the reflection component is considerable, 25\%--63\% of the total flux in Obs.~1 and 3--5.
This suggests that most of the time the disc is mainly illuminated by the corona, and the contribution of the illuminating source is not affected by the source state.
It is worthwhile to emphasize that neither the changes of {\sc cutoffpl} and the {\sc reflionx} fluxes nor these of the {\sc bbodyrad} and the {\sc reflionx\_bb} fluxes are correlated.

\subsection{Some caveats}
Note that even though compared to other models, models M2\_Cp and M2\_hd statistically give the best fits, low $\chi^2$ and null hypothesis probabilities, the iron abundance derived from these two models pegged at the upper allowed limit. If we forced the iron abundance to be 1, the fits with the reflection models, {\sc relxillCp} and {\sc relxillD}, are worse than the fit with {\sc gaussian}.
This fact may affect the other best-fitting parameters derived from both reflection models. However, the relative evolution of these parameters should be still reliable.


\section{Conclusions}
The neutron star low-mass X-ray binaries 4U~1728--34 has been jointly observed by {\xmm} and {\rx} in 2011. We carried out the spectral and timing analysis with both instruments, and found that the source evolved from the soft to the hard state during a period of $\sim40~$days. We fitted the PN spectra with several reflection models; the fits yield a disc inclination angle of $25-53\degree$ and an iron abundance as high as 10 times solar, which is probably the result of the lack of high-energy coverage of the {\xmm} instruments. Besides that, when the source evolved from the soft to intermediate state, we found that the changes in the inner radius of the accretion disc do not support the standard accretion model.
We finally concluded that during the entire evolution, both the corona and the NS surface/boundary layer contributed to the reflection component, but the former was dominant most of the time.

\section*{Acknowledgements}
This work is partly supported by China Scholarship Council (CSC), under the
grant number 201404910530. TMB acknowledges financial contribution from the agreement ASI-INAF n.2017-14-H.0. DA acknowledges support from the Royal Society. EMR acknowledges the support from Conselho Nacional de Desenvolvimento Cient\'ifico e Tecnol\'ogico (CNPq-Brazil). GZ acknowledges funding support from the CAS Pioneer Hundred Talent Program Y7CZ181001.
This work has made use of data from the High Energy Astrophysics Science Archive Research Center (HEASARC), provided by NASA/Goddard Space Flight Center (GSFC).

\bibliographystyle{mn2e}
\bibliography{ref_2018.bib}

\appendix
\section{Additional best-fitting parameters for the {\sub}} 

\begin{table*}
\caption{\label{tab:refl_hc}Best-fitting parameters for M3\_pl, {\sc tbabs*(bbodyrad+kerrconv*reflionx+cutoffpl)}}
\renewcommand{\arraystretch}{1.4}
\footnotesize
\centering
\begin{tabular}{clcccccc}
\hline \hline
\multicolumn{2}{c}{Components }& Obs.~1 &  Obs.~2 & Obs.~3 & Obs.~4 & Obs.~5 \\
\hline
{\sc tbabs} &$N\rm_{H}~(10^{22}~cm^{-2})$ &$5.3\pm0.1^{l}$&..&..&..&..\\
\hline
{\sc bbodyrad} &$T\rm_{bb}$~(keV)& $0.91\pm0.02$ &$0.12\pm0.002$ &$1.95\pm0.02$ &$1.08\pm0.03$ &$1.24_{-0.002}^{+0.03}$ \\
&$N\rm_{bb}~(km^{2}/100~kpc^2$)& $26.2\pm0.2$ &$1.5_{-0.4}^{+3.6}*10^8$ &$0.4\pm0.01$ &$4.0\pm0.1$ &$8.8_{-0.7}^{+0.04}$ \\
&flux& $1.3\pm0.01$& $0.009\pm0.002$&$0.5\pm0.01$ &$0.4\pm0.01$ & $1.8\pm0.01$\\
{\sc kerrconv} &$q\rm_{in}$ &$2.35\pm0.02^{l}$&..&..&..&..\\
& $a_{*}$ &$0.17^{f}$&..&..&..&..\\
& $i$~($\degree$) &$24.6\pm1.2^l$&..&..&..&..\\
&$R\rm_{in}$~$(R\rm_{ISCO})$&$10.7_{-0.8}^{+0.03}$ &$1.7\pm0.1$ &$1.5\pm0.1$ &$1.1_{-0.02}^{+0.3}$ &$4.6\pm0.7$ \\
{\sc reflionx}& $A\rm_{Fe}$ &$5.8_{-0.7}^{+0.04^l}$&..&..&..&..\\
&$E\rm_{cut}~(keV)$&$300^{f}$&..&..&..&..\\
&$\Gamma$&$1.58_{-0.01}^{+0.001}$ &$1.69\pm0.01$ &$1.55\pm0.004$ &$1.50_{-0.01}^{+0.002}$ &$1.4_{-0p}^{+0.02}$ \\
&$\log~\xi$ $(\rm erg~cm~s^{-1})$&$4.00_{-0.03}^{+0p}$ &$4.00\pm0.05$ &$3.84\pm0.05$ &$3.73\pm0.04$ &$3.56\pm0.06$ \\
&$N\rm_{ref}~(10^{-6})$&$1.2_{-0.03}^{+0.001}$ &$4.6\pm0.001$ &$1.0_{-0.1}^{+0.001}$ &$0.9_{-0.1}^{+0.001}$ &$1.0_{-0.001}^{+0.1}$ \\
&flux& $17.7_{-0.2}^{+0.6}$& $7.1_{-0.2}^{+0.01}$&$10.3\pm0.01$ &$6.6_{-0.01}^{+0.8}$ & $4.2_{-0.4}^{+0.02}$\\
{\sc cutoffpl} &$N\rm_{pl}$&$0.02\pm0.003$ &$0.08_{-0.01}^{+0.003}$ &$0.11_{-0.02}^{+0.002}$ &$0.14_{-0.002}^{+0.02}$ &$0.15\pm0.01$ \\
&flux& $0.9\pm0.02$& $3.0\pm0.01$&$5.5_{-0.5}^{+0.02}$ &$6.2_{-0.2}^{+0.01}$ & $9.4\pm0.01$\\
\hline
total flux&& $19.8\pm0.03$& $9.9\pm0.01$&$15.9\pm0.01$ &$12.9\pm0.01$ & $15.3\pm0.01$\\
$\chi^2/\nu$&&\multicolumn{5}{c}{$727.3/637$}\\
null hypothesis probability && \multicolumn{5}{c}{$6.8\times10^{-3}$}\\
\hline
\end{tabular}
\begin{flushleft}
{\bf Note:} In this and the following tables, the symbol $l$ indicates that the parameters are linked to vary across the observations, $f$ means that the parameter is fixed during the fit, $p$ denotes that the parameter pegs at its limit and $u$ stands for 95\% confidence upper limit. All the fluxes are in units of $10^{-10}\rm erg~cm^{-2}s^{-1}$ in the 2.5--11~keV range. Errors are quoted at $1\sigma$ confidence level.
\end{flushleft}
\end{table*}

\begin{table*}
\caption{\label{tab:refl_hd}Best-fitting parameters for M3\_hd, {\sc tbabs*(bbodyrad+kerrconv*reflionx\_hd+cutoffpl)}}
\renewcommand{\arraystretch}{1.3}
\footnotesize
\centering
\begin{tabular}{clcccccc}
\hline \hline
\multicolumn{2}{c}{Components }& Obs.~1 &  Obs.~2 & Obs.~3 & Obs.~4 & Obs.~5 \\

\hline
{\sc tbabs} &$N\rm_{H}~(10^{22}~cm^{-2})$ &$5.03\pm0.1^{l}$&..&..&..&..\\
\hline
{\sc bbodyrad} &$T\rm_{bb}$~(keV)& $2.54\pm0.002$ &$2.25\pm0.003$ &$2.55_{-0.002}^{+0.07}$ &$2.72_{-0.003}^{+0.08}$ &$3.37_{-0.1}^{+0.003}$ \\
&$N\rm_{bb}~(km^{2}/100~kpc^2$)& $2.3\pm0.01$ &$1.8\pm0.03$ &$2.3_{-0.004}^{+0.1}$ &$1.3\pm0.004$ &$0.9\pm0.002$ \\
&flux& $6.4\pm0.01$& $3.4\pm0.02$&$6.5\pm0.01$ &$4.4\pm0.01$ & $5.5\pm0.01$\\
{\sc kerrconv} &$q\rm_{in}$ &$2.32\pm0.02^{l}$&..&..&..&..\\
& $a_{*}$ &$0.17^{f}$&..&..&..&..\\
& $i$~($\degree$) &$27.6_{-0.3}^{+0.01l}$&..&..&..&..\\
&$R\rm_{in}$~$(R\rm_{ISCO})$&$397.3_{-224.7}^{+1.8}$ &$2.6\pm0.4$ &$1.6\pm0.1$ &$1.6\pm0.2$ &$5.1_{-0.3}^{+1.1}$ \\
{\sc reflionx\_hd}& $A\rm_{Fe}$ &$1^l$&..&..&..&..\\
& $\log~\rm N~(10^{22})$ &$1_{-0.05}^{+0l}$&..&..&..&..\\
&$E\rm_{cut}~(keV)$&$300^{f}$&..&..&..&..\\
&$\Gamma$&$2.27_{-0.04}^{+0.001}$ &$2.04\pm0.001$ &$2.00_{-0.01}^{+0.001}$ &$1.72_{-0.03}^{+0.001}$ &$1.4_{-0p}^{+0.002}$ \\
&$\log~\xi$ $(\rm erg~cm~s^{-1})$&$2.97\pm0.09$ &$2.83\pm0.09$ &$2.89\pm0.08$ &$2.97\pm0.08$ &$3.08\pm0.05$ \\
&$N\rm_{ref}$&$1.4\pm0.001$ &$0.3_{-0.01}^{+0.001}$ &$0.7\pm0.001$ &$0.4\pm0.001$ &$0.5\pm0.001$ \\
&flux& $10.8_{-0.01}^{+1.2}$& $3.1_{-0.1}^{+0.01}$&$7.1\pm0.01$ &$6.2_{-1.0}^{+0.1}$ & $9.3_{-0.8}^{+0.01}$\\
{\sc cutoffpl} &$N\rm_{pl}$&$0.16\pm0.002$ &$0.15\pm0.02$ &$0.11_{-0.04}^{+0.001}$ &$0.07_{-0.008}^{+0.002}$ &$0.007\pm0.002$ \\
&flux& $2.4\pm0.02$& $3.3_{-0.2}^{+0.01}$&$2.5\pm0.01$ &$2.4_{-0.4}^{+0.01}$ & $0.5_{-0.2}^{+0.01}$\\
\hline
total flux&& $19.6\pm0.01$& $9.8\pm0.01$&$15.7\pm0.01$ &$12.8\pm0.01$ & $15.1\pm0.03$\\
$\chi^2/\nu$&&\multicolumn{5}{c}{$763.5/637$}\\
null hypothesis probability && \multicolumn{5}{c}{$3.7\times10^{-4}$}\\
\hline
\end{tabular}
\begin{flushleft}
{\bf Note:} All the symbols and units are the same as in Table~A1. 
\end{flushleft}
\end{table*}

\begin{table*}
\caption{\label{tab:refl_bb}Best-fitting parameters for M3\_bb, {\sc tbabs*(bbodyrad+kerrconv*reflionx\_bb+cutoffpl)}}
\renewcommand{\arraystretch}{1.3}
\footnotesize
\centering
\begin{tabular}{clcccccc}
\hline \hline
\multicolumn{2}{c}{Components }& Obs.~1 &  Obs.~2 & Obs.~3 & Obs.~4 & Obs.~5 \\

\hline
{\sc tbabs} &$N\rm_{H}~(10^{22}~cm^{-2})$ &$6.44_{-0.07}^{+0.01}$&..&..&..&..\\
\hline
{\sc bbodyrad} &$T\rm_{bb}$~(keV)& $2.02\pm0.002$ &$2.01\pm0.003$ &$2.06\pm0.01$ &$1.97\pm0.01$ &$2.01\pm0.03$ \\
&$N\rm_{bb}~(km^{2}/100~kpc^2$)& $4.3_{-0.2}^{+0.01}$ &$2.2_{-0.07}^{+0.2}$ &$2.6\pm0.01$ &$0.9_{-0.2}^{+0.08}$ &$1.2_{-0.2}^{+0.01}$ \\
&flux& $5.2_{-0.01}^{+0.3}$& $2.9_{-0.09}^{+0.01}$&$3.7\pm0.3$ &$1.1\pm0.4$ & $1.5\pm0.2$\\
{\sc kerrconv} &$q\rm_{in}$ &$3.84\pm0.16^{l}$&..&..&..&..\\
& $a_{*}$ &$0.17^{f}$&..&..&..&..\\
& $i$~($\degree$) &$52.9_{-0.5}^{+1.6l}$&..&..&..&..\\
&$R\rm_{in}$~$(R\rm_{ISCO})$&$7.1_{-0.4}^{+0.002}$ &$8.7\pm1.3$ &$6.2\pm0.5$ &$5.9\pm0.5$ &$9.3\pm0.7$ \\
{\sc reflionx\_bb}& $A\rm_{Fe}$ &$0.78_{-0.09}^{+0.01l}$&..&..&..&..\\
&$E\rm_{cut}~(keV)$&$300^{f}$&..&..&..&..\\
&$\log~\xi$ $(\rm erg~cm~s^{-1})$&$2.20\pm0.002$ &$1.85_{-0.04}^{+0.003}$ &$1.80_{-0.002}^{+0.02}$ &$1.81_{-0.03}^{+0.002}$ &$2.05_{-0.05}^{+0.002}$ \\
&$N\rm_{ref}$&$1.1\pm0.2$ &$2.2\pm0.02$ &$5.2\pm0.4$ &$4.7\pm0.4$ &$1.6\pm0.01$ \\
&flux& $2.2_{-0.1}^{+0.01}$& $1.5_{-0.1}^{+0.01}$&$3.0_{-0.2}^{+0.01}$ &$2.9_{-0.08}^{-0.01}$ & $2.1\pm0.01$\\
{\sc cutoffpl} &$\Gamma$&$2.62_{-0.03}^{+0.001}$ &$2.59_{-0.001}^{-0.04}$ &$2.42_{-0.01}^{+0.001}$ &$2.12_{-0.001}^{+0.02}$ &$1.89_{-0.01}^{+0.001}$ \\
&$N\rm_{pl}$&$1.48\pm0.04$ &$0.65\pm0.001$ &$0.83_{-0.001}^{+0.02}$ &$0.50_{-0.02}^{+0.001}$ &$0.44_{-0.0004}^{+0.006}$ \\
&flux& $12.7_{-0.2}^{+0.07}$& $5.7\pm0.1$&$9.6\pm0.2$ &$9.4\pm0.01$ & $12.3\pm0.2$\\
\hline
total flux& &$20.8\pm0.1$& $10.3\pm0.02$&$16.6\pm0.03$ &$13.5\pm0.04$ & $16.0\pm0.03$\\
$\chi^2/\nu$&&\multicolumn{5}{c}{$753.2/637$}\\
null hypothesis probability && \multicolumn{5}{c}{$3.9\times10^{-5}$}\\
\hline
\end{tabular}
\begin{flushleft}
{\bf Note:} All the symbols and units are the same as in Table~A1. 
\end{flushleft}
\end{table*}

\begin{table*}
\caption{\label{tab:refl_pl_bb}Best-fitting parameters for M3\_pl\_bb, {\sc tbabs*(bbodyrad+kerrconv*(reflionx+reflionx\_bb)+cutoffpl)}}
\renewcommand{\arraystretch}{1.3}
\footnotesize
\centering
\begin{tabular}{clcccccc}
\hline \hline
\multicolumn{2}{c}{Components }& Obs.~1 &  Obs.~2 & Obs.~3 & Obs.~4 & Obs.~5 \\

\hline
{\sc tbabs} &$N\rm_{H}~(10^{22}~cm^{-2})$ &$5.20\pm0.01^{l}$&..&..&..&..\\
\hline
{\sc bbodyrad} &$T\rm_{bb}$~(keV)& $1.04_{-0.003}^{+0.03}$ &$2.71_{-0.07}^{+0.01}$ &$1.95_{-0.03}^{+1.38}$ &$1.15_{-0.04}^{+0.006}$ &$1.25\pm0.002$ \\
&$N\rm_{bb}~(km^{2}/100~kpc^2$)& $5.0\pm0.3$ &$0.04_{-0.004}^{+0.01}$ &$0.1\pm0.01$ &$0.5\pm0.06$ &$9.1_{-0.1}^{+0.04}$ \\
&flux& $0.5\pm0.03$& $0.1_{-0.1}^{+0.01}$&$0.1_{-0.1}^{+0.01}$ &$0.08_{-0.08}^{+0.01}$ & $1.9\pm0.02$\\
{\sc kerrconv} &$q\rm_{in}$ &$2.49\pm0.06^{l}$&..&..&..&..\\
& $a_{*}$ &$0.17^{f}$&..&..&..&..\\
& $i$~($\degree$) &$27.7_{-0.6}^{+0.07l}$&..&..&..&..\\
&$R\rm_{in}$~$(R\rm_{ISCO})$&$56.7_{-1.0}^{+0.3}$ &$2.5\pm0.3$ &$1.7\pm0.1$ &$1.5\pm0.2$ &$4.3\pm0.6$ \\
{\sc reflionx(\_bb)}& $A\rm_{Fe}$ &$6.76_{-0.21}^{+0.04l}$&..&..&..&..\\
&$E\rm_{cut}~(keV)$&$300^{f}$&..&..&..&..\\
&$\Gamma$&$1.47\pm0.02$ &$2.20\pm0.02$ &$1.54_{-0.004}^{+0.02}$ &$1.41_{-0.002}^{+0.02}$ &$1.40\pm0.001$ \\
&$\log~\xi$ $(\rm erg~cm~s^{-1})$&$4.00_{-0.16}^{+0.05}$ &$3.92\pm0.07$ &$3.86\pm0.06$ &$3.75\pm0.05$ &$3.55_{-0.16}^{+0.15}$ \\
&$N\rm_{ref\_pl}~(10^{-6})$&$1.2\pm0.003$ &$<0.02^{u}$ &$0.9\pm0.003$ &$0.9\pm0.003$ &$0.9\pm0.01$ \\
&flux& $15.7\pm0.1$& $<0.1^{u}$&$9.8_{-0.1}^{+0.01}$ &$6.1\pm0.01$ & $3.8\pm0.01$\\
&$N\rm_{ref\_bb}$&$0.07\pm0.001$ &$0.1\pm0.003$ &$0.01\pm0.004$ &$0.05\pm0.005$ &$<0.006^{u}$ \\
&flux& $1.7_{-1.0}^{+0.1}$& $4.2\pm0.3$&$0.7_{-0.3}^{+3.4}$ &$1.0_{-0.6}^{+1.4}$ & $<0.05^{u}$\\
{\sc cutoffpl} &$N\rm_{pl}$&$0.03\pm0.003$ &$0.35_{-0.001}^{+0.02}$ &$0.12\pm0.01$ &$0.10\pm0.0002$ &$0.15_{-0.001}^{+0.0001}$ \\
&flux& $1.6\pm0.2$& $5.9_{-0.3}^{+0.04}$&$6.0\pm0.3$ &$6.1\pm0.02$ & $9.6_{-0.01}^{+0.4}$\\
\hline
total flux&& $19.7\pm0.03$& $9.8\pm0.03$&$15.8\pm0.1$ &$12.9\pm0.03$ & $15.2\pm0.02$\\
$\chi^2/\nu$&&\multicolumn{5}{c}{$730.2/634$}\\
null hypothesis probability && \multicolumn{5}{c}{$3.2\times10^{-3}$}\\
\hline
\end{tabular}
\begin{flushleft}
{\bf Note:} All the symbols and units are the same as in Table~A1. 
\end{flushleft}
\end{table*}
\end{document}